\documentclass[%
reprint,
superscriptaddress,
notitlepage,
amsmath,amssymb,
aps,
onecolumn,
floatfix,
tightenlines,
longbibliography,
11pt,
]{revtex4-1}

\usepackage[usenames,dvipsnames,svgnames,table]{xcolor}
\usepackage{psfrag,graphicx,epsfig,color}% Include figure files
\usepackage{dcolumn}% Align table columns on decimal point
\usepackage{bm}% bold math
\usepackage{natbib}
\usepackage{float}
\usepackage{subfigure}

\usepackage{tikz}
\usetikzlibrary{shapes}
\usetikzlibrary{positioning}
\usetikzlibrary{calc,3d}

\definecolor{apricot}{rgb}{0.98, 0.81, 0.69}
\definecolor{bisque}{rgb}{1.0, 0.89, 0.77}

\newcommand\ve[1]{\boldsymbol{#1}}

\def\ww {{\boldsymbol{\omega}}}

\def\xx {{\mathbf{x}}}

\begin{document}

\title{Cascades and Reconnection in Interacting Vortex Filaments}
%\title{Reconnection of Interacting Vortex Tubes and Cascades}

\author{Rodolfo Ostilla-M\'{o}nico}
\affiliation{Department of Mechanical Engineering, University of Houston, Houston, TX 77204, USA}
\author{Ryan McKeown}
\affiliation{School of Engineering and Applied Sciences, Harvard University, Cambridge, MA 02138, USA}

\author{Michael P. Brenner}
\affiliation{School of Engineering and Applied Sciences, Harvard University, Cambridge, MA 02138, USA}
\author{Shmuel M. Rubinstein}
\affiliation{School of Engineering and Applied Sciences, Harvard University, Cambridge, MA 02138, USA}
\author{Alain Pumir}
\affiliation{Universit\'e de Lyon, ENS de Lyon, Universit\'e Claude Bernard, CNRS, Laboratoire de Physique, 69342 Lyon, France}
\affiliation{Max Planck Institute for Dynamics and Self-Organization, 37077 G\"ottingen, Germany}

\date{\today}

\begin{abstract}
At high Reynolds number, 
the interaction between two vortex tubes leads to intense velocity gradients, which are at the heart of fluid turbulence. This vorticity amplification comes about through  two different instability mechanisms of the initial vortex tubes, assumed anti-parallel and with a mirror plane of symmetry. At moderate Reynolds number, the tubes destabilize via a Crow instability, with the nonlinear development leading to strong flattening of the cores into thin sheets. These sheets then break down into filaments which can repeat the process. At higher Reynolds number, the instability proceeds via the elliptical instability, producing vortex tubes that are perpendicular to the original tube directions.
In this work, we demonstrate that these same transition between Crow and Elliptical instability occurs at moderate Reynolds number when we vary the initial {\sl angle} $\beta$ between two straight vortex tubes. 
We demonstrate that when the angle between the two tubes is close to $\pi/2$, the interaction between tubes leads to the formation 
of thin vortex sheets. The subsequent breakdown 
of these sheets involves a twisting of the paired sheets, followed
by the appearance of a localized cloud of small scale vortex structures. 
At smaller values of the angle $\beta$ between the two tubes, the breakdown mechanism changes to an elliptic
cascade-like mechanism. Whereas 
the interaction of two vortices depends on the initial condition, the rapid formation of fine-scales vortex structures appears to 
be a robust feature, possibly universal at very high Reynolds numbers.
\end{abstract}

\maketitle

\section{introduction}
\label{sec:intro}

Many experiments have demonstrated that the interaction of 
two vortex tubes coming close together eventually leads to 
a change of topology of the vortex lines through a process known as 
vortex reconnection~\cite{Saffman:1981,Schatzle:1987,Saffman:1992}. 
Vortex reconnection is a fundamental process in fluid mechanics, 
and it has been postulated to play a significant role in fluid
phenomena such as the turbulent energy cascade \cite{Goto:2012}, noise 
generation \cite{Hussain:1986} and the transfer of helicity across
topologically distinct vortices \cite{Scheeler:2017}.
Reconnection is also interesting from a theoretical point of view, as
the change of vortex line topology appears to violate well-known 
conservation theorems in inviscid flows~\cite{Batchelor:1970}, which 
implies that viscosity must play a decisive role even at extreme Reynolds numbers.

The process through which vortex pairs undergo reconnection
has been well studied and characterized both numerically and theoretically.
The early phase of the interaction, before 
viscosity plays the dominant role, is captured
by the Biot-Savart equation, which keeps track of the location of
the vortex tubes by assuming a  circular structure of the 
cores, with a fixed vorticity profile~\cite{Leonard:1980}. The
numerical work of~\cite{Siggia:1985} showed that 
the resulting dynamics of a wide range of initial conditions
spontaneously leads to the local pairing of antiparallel parts of nearby filaments. As a result,
the cores get close together, which generally leads to very rich dynamics.
The Biot-Savart description, however, fails when the vortex tubes are so close 
that they deform each others' cores, thereby making the initial assumption 
questionable~\cite{Ashurst:1987,Pumir:1987}.
Therefore, to adequately capture the initial dynamics of the interaction, it is necessary 
to undertake a full simulation of the Euler equations. 
The Biot-Savart equation (and the Euler equations) also fail to capture
the viscous processes, essential to reconnection, which occur at scales 
much smaller than any other inviscid process scale~\cite{Pumir:1987}. 

It has been known for a long time that several instability mechanisms
may lead to the disruption of two antiparallel vortex lines. 
Early studies based on the Biot-Savart equation suggested that the 
long-wavelength Crow-instability~\cite{Crow:1970} plays the dominant role in
bringing together counter-rotating parts of the 
tubes so reconnection can occur~\cite{SiggiaP:1985,PumirSig:1987}. This prompted a number of 
investigations, see e.g.\cite{Hussain:1986,Pumir:1990,Shelley:1993,Kerr:1993,Hou:2006,Brenner:2016,Yao:2020a}, based on numerical solutions of the Euler and Navier-Stokes
equations which imposed the symmetry of the most
unstable mode corresponding to the long wave-number Crow instability, namely that of 
two tubes with two mirror symmetry in the plane that separates them (denoted as
the plane $P_1$).
With this symmetry, the components of velocity perpendicular to the plane $P_1$
is uniformly zero, so if the vorticity component perpendicular to the plane is zero,
then, reconnection is impossible in the absence of viscosity.
The observation that the time necessary to achieve reconnection, in the limit 
of very small viscosity, $\nu$, seems in practice to be independent of viscosity, 
suggests that
the limit $\nu \to 0$~\cite{Pumir:1987} (or alternatively, of the limit $Re_\Gamma\to\infty$, where $Re_\Gamma$, the Reynolds number is defined as $Re_\Gamma = \Gamma/\nu$, $\Gamma$ being the circulation of the vortex tubes and $\nu$ the kinematic viscosity) is singular. 
This observation has been interpreted as
a signature pointing to the existence of singular solutions 
of the Euler equations. The study of simplified models, based on
the Biot-Savart equations, has suggested a large amplification of vorticity during the late stages of reconnection, although
the approximations necessary to derive the Biot-Savart model ultimately break 
down~\cite{SiggiaP:1985,Hormoz:2012,Moffatt:2019}. Furthermore,
decades of careful numerical work have not conclusively resolved 
the singularity issue for the corresponding initial value problem. 
The problem has been particularly studied in 
the inviscid limit, looking for signs of a diverging 
vorticity~\cite{Pumir:1990,Kerr:1993,Hou:2006}. Still, DNS clearly show that very thin vortex sheets are formed 
on either side of the symmetry plane $P_1$, both in the inviscid and in the 
viscous problem at large enough $Re_\Gamma$. The formation of extremely thin vortex sheets, with 
a relatively slow growth of vorticity makes the problem very difficult
to study numerically. 

While the quest to reach increasingly large values of $Re_\Gamma$ continues \cite{van_Rees:2012}, 
numerical and experimental studies have recently identified another
main mechanism in the interaction between two antiparallel tubes, which leads to 
the breaking of the vortex tubes instead of to a reconnection \cite{McKeown:2020}.
This mechanism can be observed in the head-on collision 
between two vortex rings, which leads to a very rapid destruction of 
the vortices at large Reynolds numbers~\cite{Lim:1992}. In this problem, the
long-wavelength Crow instability initially brings parts of the filaments 
together~\cite{Leweke:2016,McKeown:2018}. However, the further interaction between the counter-rotating tubes,
reveals the dominant role of an instability whose wavenumber
is comparable with the core size~\cite{Tsai:1976,Moore:1975}.
This instability, known as the 
elliptic instability~\cite{Bayly:1986,Kerswell:2002,Leweke:2016}, involves
a symmetry that completely differs from that of the long-wavenumber
Crow instability.
At sufficiently high Reynolds numbers, elliptic instabilities develop on top of each
other, leading to a cascade, and eventually, to a transient turbulent flow~\cite{McKeown:2020}.

The marked difference between these two cases leads us to investigate the onset of the mechanisms leading 
to reconnection. In this vein, we take inspiration from 
the studies of reconnecting magnetic tubes at an angle in an 
astrophysical context~\cite{Linton:2001}, and from 
the recent study of vortex reconnection in superfluids, which has
revealed the presence of non-universal features by comparing different classes
of initial conditions~\cite{Villois:2017}. However, we have
to highlight the notable difference between
these two cases, and ours, which arises from much larger degrees of freedom of the vortex
cores in hydrodynamics and leads to a much richer phenomenology, not taken 
into account in the simplified model of reconnection of skewed vortices in~\cite{Kimura:1994}.
While it is clear that reconnections are still present
in hydrodynamical fluids with $Re_\Gamma\gg 1$ \cite{van_Rees:2012}, their appearance could be restricted to a rather small subset of initial conditions that either enforce many symmetries on the vortices or  
stabilize the core through spin \cite{Yao:2020a}. This would mean that reconnections (understood as topological changes) become rare in a classical 
fluid and play a small role in the conveyance
of energy across scales, even if they remain important from a theoretical and mathematical perspective~\cite{Fefferman:2006}. 

We address the question about genericity by relaxing some of the strict symmetries imposed
on the usual reconnection studies. We perform a series of 
direct numerical simulations which use  the simple configuration of 
two counter-rotating vortex tubes oriented at an angle as the initial condition. We work at a Reynolds number where for antiparallel tubes, the elliptical 
instability dominates, due to the amplification of strain that results from their alignment. We start off with
two filaments which are perpendicular to each other. For this configuration, we will see that the strain orientation does not  
excite the elliptical instability. 
We then vary the angle, making the vortices more antiparallel and
 recover the elliptical instability. 
In practice, we use $ b = \cot(\beta/2)$, where $\beta$ is the angle 
between the filaments. With this choice, $b = 1$ corresponds to two initially
perpendicular filaments: $\beta = \pi/2$.
In this study, we consider only values of 
$\beta \le \pi/2$  ($b \ge 1$), thus favoring configurations
where the two filaments tend to be initially counter-rotating, rather than co-rotating.
We note that the configuration $\beta = \pi/2$ ($b = 1$) has been studied, originally at much smaller Reynolds numbers 
than the ones considered here and with an additional
hyperviscous dissipation term~\cite{Boratav:1992}, and more recently at a much higher resolution~\cite{Yao:2020c}. 

In all cases, we find an energy cascade  during the interaction, reaching ever smaller scales as the Reynolds number increases. This
generation of small scales arises from deformations of the cores 
where the tubes intersect. We find that the interaction starts with the formation of characteristic 
vortex sheets for $67.4^\circ\le\beta\le90^\circ$ ($1 \le b \le 3/2$). For $\beta\le53.1^\circ$ ($b \ge 2$), however, the mechanism that 
prevails in the interaction between the two tubes is the formation of 
transverse vortex tubes, as observed, formally, when $\beta \to 0$ ($b \to \infty$) due to the presence of the elliptical instability \cite{McKeown:2020}.

\section{Numerical procedures and database}
\label{sec:num}

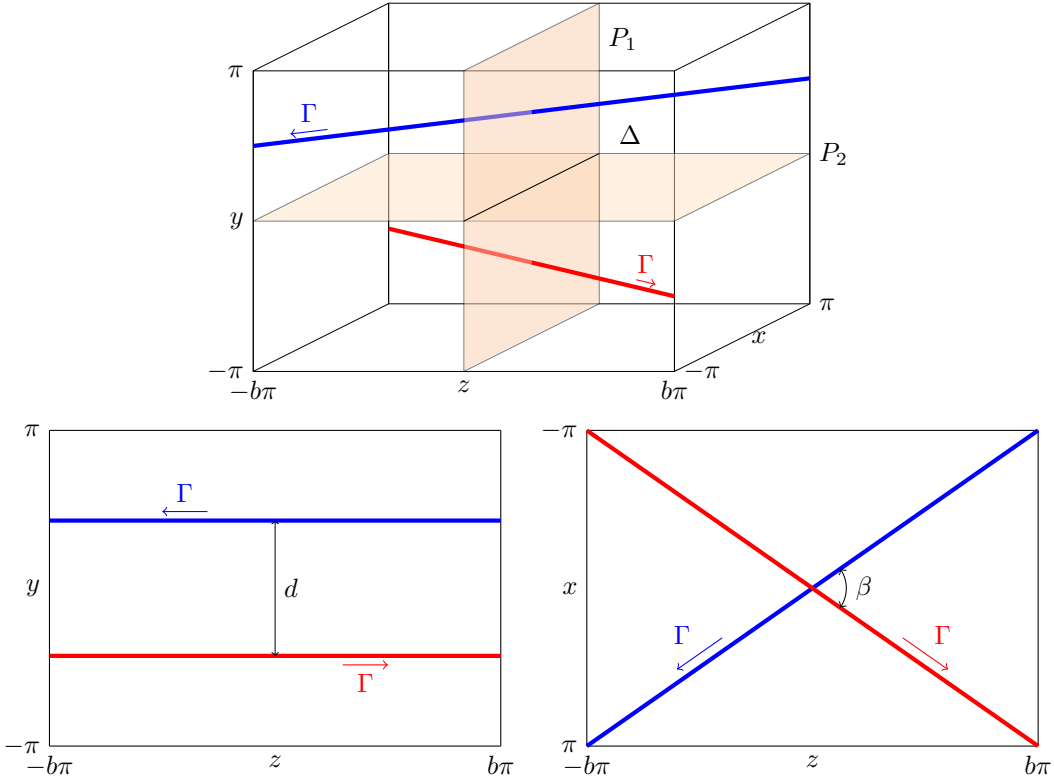
\begin{figure}[tb]
\begin{center}
\begin{tikzpicture}[x  = {(-0.3cm,-0.15cm)},
                    y  = {(0.7cm,0cm)},
                    z  = {(0cm,1cm)},
                    scale = 2]
% style of faces
\tikzset{facestyle/.style={fill=none,draw=black,very thin,line join=round}}
% face "back" 
\begin{scope}[canvas is zy plane at x=0]
  \path[facestyle] (0,0) rectangle (2,4);
\end{scope}
% face  "left"
\begin{scope}[canvas is zx plane at y=0]
  \path[facestyle] (0,0) rectangle (2,3);
\end{scope}
% face "front"
\begin{scope}[canvas is zy plane at x=3]
  \path[facestyle] (0,0) rectangle (2,4);
\end{scope}
% face  "right"
\begin{scope}[canvas is zx plane at y=4]
  \path[facestyle] (0,0) rectangle (2,3);
\end{scope}

\draw[red, ultra thick] (0,0,0.5) --  (1.5,2,0.5);
\draw[->][red, thin](2.6,3.466,0.55)-- node[above] {$\Gamma$} (2.8,3.733,0.55);

\draw[blue, ultra thick] (1.5,2,1.5) --  (3,0,1.5); 
\draw[->][blue, thin](2.6,0.534,1.55)-- node[above] {$\Gamma$} (2.8,0.267,1.55);
%\draw[black, ultra thin] (1.5,2,0.5) --  (1.5,2,1.5);

% face "up" 
\draw[fill=bisque,draw=black,opacity=.5,very thin,line join=round]
 (0,0,1) --  (3,0,1) --  (3,4,1) -- (0,4,1) --cycle ;
 
\draw[fill=apricot,draw=black,opacity=.5,very thin,line join=round]
 (3,2,0) --  (3,2,2) --  (0,2,2) -- (0,2,0) --cycle ;
% labels

\draw[red, ultra thick] (1.5,2,0.5) --  (3,4,0.5);
\draw[blue, ultra thick] (0,4,1.5) --  (1.5,2,1.5);

\fill[black](1.5,4,0) node [sloped, right]{$x$};
\fill[black](0,4,0) node [sloped, right]{$\pi$};
\fill[black](3,4,0) node [sloped, right]{$-\pi$};

\fill[black](3,2,0) node [sloped, below]{$z$};
\fill[black](3,0,0) node [sloped, below]{$-b\pi$};
\fill[black](3,4,0) node [sloped, below]{$b\pi$};

\fill[black](3,0,1) node [sloped, left]{$y$};
\fill[black](3,0,0) node [sloped, left]{$-\pi$};
\fill[black](3,0,2) node [sloped, left]{$\pi$};

\draw[black] (0,2,1) --  (3,2,1);
\fill[black](0,2.1,1) node [sloped, above right]{$\Delta$};

\fill[black](0,2,1.9) node [sloped, below right]{$P_1$};
\fill[black](0,4,1) node [sloped, right]{$P_2$};
\end{tikzpicture}

\begin{tikzpicture}[scale=0.6]

\draw[black](0,7)-- (10,7);
\draw[black](0,0)-- (0,7);
\draw[black](10,0)-- (10,7);
\draw[black](0,0)-- (10,0);

\draw[blue, ultra thick](0,5)-- (10,5);
\draw[<-][blue, thin](2.5,5.2)-- node[above] {$\Gamma$} (3.5,5.2);

\draw[red, ultra thick](0,2)-- (10,2);
\draw[<-][red, thin](7.5,1.8)-- node[below] {$\Gamma$} (6.5,1.8);

\fill[black](5,0) node [scale=1,anchor= north]{$z$};
\fill[black](10,0) node [scale=1,anchor= north]{$b\pi$};
\fill[black](0,0) node [scale=1,anchor= north]{$-b\pi$};

\fill[black](0,3.5) node [scale=1,anchor= east]{$y$};
\fill[black](0,0) node [scale=1,anchor= east]{$-\pi$};
\fill[black](0,7) node [scale=1,anchor= east]{$\pi$};

\draw[<->][black, ultra thin] (5,2) -- node[right]{$d$} (5,5);

\end{tikzpicture}
\begin{tikzpicture}[scale=0.6]

\draw[black](0,7)-- (10,7);
\draw[black](0,0)-- (0,7);
\draw[black](10,0)-- (10,7);
\draw[black](0,0)-- (10,0);

\draw[blue, ultra thick](0,0)-- (10,7);
\draw[<-][blue, thin](2,1.7)-- node [above left] {$\Gamma$} (3,2.4);

\draw[red, ultra thick](10,0)-- (0,7);
\draw[->][red, thin](7,2.4)-- node [above right] {$\Gamma$} (8,1.7);

\fill[black](5,0) node [scale=1,anchor= north]{$z$};
\fill[black](10,0) node [scale=1,anchor= north]{$b\pi$};
\fill[black](0,0) node [scale=1,anchor= north]{$-b\pi$};

\fill[black](0,3.5) node [scale=1,anchor= east]{$x$};
\fill[black](0,0) node [scale=1,anchor= east]{$\pi$};
\fill[black](0,7) node [scale=1,anchor= east]{$-\pi$};

\draw[<->][black,domain=-35:35] plot ({5+0.75*cos(\x)}, {3.5+0.75*sin(\x)});
\fill[black](5.75,3.5) node [scale=1,anchor= west]{$\beta$};
\end{tikzpicture}
\caption{ Schematic of DNS initial configuration. The thick red and blue lines represent the initial positions of the two vortex filaments, with the arrow indicating the circulation direction. (a) 3D schematic. 
(b) side view of the $yz$-plane, which shows the definition of the spacing between the filaments, $d$.
(c) top view of the $xz$-plane, which shows the definition of $\beta$ and $b$.  
The vorticity distribution is invariant under the symmetry resulting from 
two mirror symmetries with respect to the planes $P_1$ and $P_2$:
$(x,y,z) \rightarrow (x, -y, -z) $;
$(\omega_x, \omega_y , \omega_z) \rightarrow (\omega_x, -\omega_y, -\omega_z) $
and $(u_x, u_y , u_z) \rightarrow (u_x, -u_y, -u_z) $.
The distance between the two tubes in the $y$ direction is $d = 0.9$; the
circulation is chosen here to be $\Gamma = 1$, and the core radius 
$\sigma = 0.4/\sqrt{2} \approx 0.28$.
}
\label{fig:scheme}
\end{center}
\end{figure}

 We simulate the incompressible Navier-Stokes equations:
\begin{eqnarray}
& & \partial_t {\ve u} + ( \ve u \cdot \nabla ) \ve u = - \rho^{-1} \nabla p + \nu \nabla^2 \ve u \label{eq:NS} \\
& & \nabla \cdot {\ve u}  =  0 \label{eq:incomp}
\end{eqnarray}
in a triply periodic box, using pseudo-spectral methods. 
The details of the code have been described in~\cite{Pumir:1994}. 
We vary the aspect ratio of the domain, which we take to be of size $2 \pi$ in
the $x$ and $y$ direction, and of size $2 b \pi$ in the $z$ direction,
where $b$ is a control parameter
that we take as $b=1$, $5/4$, $3/2$, $2$, $5/2$, $3$, and $4$.
No forcing is added to the Navier-Stokes equations, and the flow is
allowed to evolve from the initial conditions.
These consist of two Gaussian vortices, where the initial position of the
vortex cores are in two diagonal lines $\gamma_\pm$ $ z = \pm b x$ located at the plane $y = \pm d/2$ . The vorticity is initially
concentrated around the two lines, with a Gaussian distribution:
$\ww_\pm( \ve x, t = 0) = 
\pm \Omega \exp( - \ve \rho_\pm^2/2 \sigma^2 ) 
( \ve e_x \pm b \ve e_z)/\sqrt{1 + b^2} $, 
where $ \ve \rho_\pm$ is the distance between the point $\ve x$ 
to the two lines $\gamma_\pm$, and $\sigma$ is the core radius. 
The resulting circulations, $\Gamma_\pm$, are equal to 
$\Omega \sigma^2$.
A schematic of the initial condition can be seen in Fig.~\ref{fig:scheme}, 
which shows that the iso-contours of vorticity approximately
concentrate in two tubes, at an angle of inclination $\beta = 2 \, \arctan(1/b)$.

Our calculations are organized in two series of runs. In the first series, 
we fix the angle by setting $b = 1$, which results in $\beta=90^\circ$ to 
each other. This configuration minimizes the strain direction that triggers the 
elliptical instability. We then vary the Reynolds number, from 
$Re_{\Gamma} = 2200$ to $Re_{\Gamma} = 5400$, which is around the Reynolds number range for which the elliptical instability
supersedes the Crow instability for antiparallel tubes, to study the genericity of the elliptical 
instability in the most disadvantageous configuration.
In the second series of runs, we fix the Reynolds number at $Re_{\Gamma} = 4000$, and vary 
$b$ from $1$ to $4$, which reduces the angle, $\beta$, and brings the tubes closer to being
antiparallel, progressively amplifying the strain in the direction that induces the elliptical
instability.

In addition, we considered 3 runs with initially antiparallel vortex tubes, 
in the configuration studied in~\cite{McKeown:2018,McKeown:2020}. These runs were carried out 
in a box of aspect ratio 4, although they correspond formally to $\beta = 0$, hence
$b \rightarrow \infty$. In these runs, we kept the Reynolds number to $Re_{\Gamma} = 4000$, and we 
slightly modulated the constant $x$ and $y$-locations of the tubes by a
sum of a few Fourier modes. We varied the overall coefficient of the perturbation by multiplying
by $2$ and $4$. The runs were carried out at low resolution ($192^2 \times 768$). Complementary
runs at higher resolution convinced us that the low resolution was in fact sufficient.

In all the runs, the initial evolution is relatively smooth and does not
require a very high resolution. We have therefore started all the runs at
a low resolution, with a grid of size $N_l \times N_l \times (b \, N_l)$
(or equivalently, with as many Fourier modes).
When the vortex tubes come together, the velocity 
field develops very fine scales, or equivalently, the Fourier spectrum extends
to much larger values of the wavenumbers, $k$. To simulate this phase of the
dynamics, we extend the number of Fourier modes to 
$N_h \times N_h \times (N_h \, b)$.
The parameters of the various simulations are shown in Table~\ref{tab:1}.

%%%%%%%%%%%%%%%%%%%%%%%%%%%%%%%%%%%%%%%%%%%%%%%%%%%%%%%%%%%%%%%
%% Table 1, DNS Parameters

\begin{table}[tb]
\centering
    \begin{tabular}{c|cccccccccccc}
 {\rm Runs} & $1$ & $2$ & $3$ & $4$ & $5$ & $6$ & $7$ & $8$ & $9$ & $10$ & $11$ &$12A/B/C$ \\
\hline
  $b$ & $1$ & $1$ & $1$ & $1$ & $1$ & $5/4$ & $3/2$ & $2$ & $5/2$ & $3$ & $4$ & $\infty$ \\
  $\beta$ & $90^\circ$ & $90^\circ$ & $90^\circ$ & $90^\circ$ & $90^\circ$ & $77.3^\circ$ & $67.4^\circ$ & $53.1^\circ$ & $43.6^\circ$ & $36.8^\circ$ & $28.1^\circ$ & $0^\circ$ \\
  $Re_\Gamma$ & $2200$ & $3300$ & $4000$ & $4550$ & $5400$ & $4000$ & $4000$ & $4000$ & $4000$ & $4000$ & $4000$ & $4000$ \\
  $N_l$ & $256$ & $256$ & $384$ & $384$ & $384$ & $192$ & $240$ & $192$ & $192$ & $192$ & $192$ & $192^*$ \\
  $N_h$ & $384$ & $512$ & $512$ & $512$ & $512$ & $384$ & $400$ & $320$ & $320$ & $320$ & $320$ & $320^*$ \\
    \end{tabular}
\caption{Simulation parameters for the runs
used in this work. 
The numerical domain is taken as $ -\pi \le x \le \pi$, $-\pi \le y \le \pi$
and $-b \pi \le z \le b \pi$. The Reynolds number is defined as the ratio
of the initial circulation, $\Gamma$, divided by the kinematic viscosity, $\nu$.
Each run was started at low resolution, 
with $N_l \times N_l \times b N_l$ Fourier modes. 
The runs were also conducted at a higher resolution, 
with $N_h \times N_h \times b N_h$ Fourier modes during the generation of small-scale flow structures. $~^*$For these cases with
$\beta = 0^\circ$, the limits in the $z$ direction are taken as $-4 \pi \le z\le 4\pi$, and the $z$ resolution is accordingly $4N_l$ or $4N_h$.}
\label{tab:1}
\end{table}
%%%%%%%%%%%%%%%%%%%%%%%%%%%%%%%%%%%%%%%%%%%%%%%%%%%%%%%%%%%%%%%

%%%%%%%%%%%%%%%%%%%%%%%%%%%%%%%%%%%%%%%%%%%%%%%%%%%%%%%%%%%%%%%
%% Figure 1 Schematic of DNS Initial Configuration

%%%%%%%%%%%%%%%%%%%%%%%%%%%%%%%%%%%%%%%%%%%%%%%%%%%%%%%%%%%%%%%

In the following, all quantities will be expressed in units defined with the box-size and a unitary circulation. Note that the time scale associated with the inviscid evolution (Biot-Savart model) is $\sim d^2/\Gamma$, which is of order $1$.

\paragraph*{Symmetries of the problem:}
Although the planes $P_1$ ($z = 0$) and $P_2$ ($y = 0$), as
indicated in Fig.~\ref{fig:scheme}, play a particularly important role
in the problem studied here, the velocity and vorticity fields in 
our simulations do not have any simple symmetry 
with respect to $P_1$ or $P_2$. In configurations with a 
symmetry with respect to $P_1$, as it is the case e.g. 
in ~\cite{Pumir:1987,Kerr:1989,Pumir:1990,Kerr:1993,Shelley:1993,Hou:2006,Kerr:2013,Brenner:2016,Yao:2020a}, 
the component of the velocity field
perpendicular to $P_1$ is equal to $0$ in $P_1$. As 
a consequence of this symmetry, 
if the component of vorticity perpendicular to $P_1$ is 
initially $0$ in the symmetry plane $P_1$, then, a component of vorticity 
perpendicular to $P_1$ cannot be generated without viscosity. 

Conversely, there is no particular symmetry plane between two
vortex tubes undergoing the elliptic instability~\cite{McKeown:2018,McKeown:2020},
or in the configuration of two vortex tubes initially at a finite angle. 
Nonetheless,
the fields in the present
study are invariant after composing the two mirror symmetries with 
respect to $P_1$ and to $P_2$, or equivalently, by a rotation with 
respect their intersection, i.e. the straight line $\Delta$ ($z = y = 0$) shown in Fig.~\ref{fig:scheme}(a).
This corresponds to the following symmetry:
\begin{equation}
(x,y,z ) \rightarrow ( x, -y, -z)  ,~~~ (u_x,u_y,u_z) \rightarrow (u_x, -u_y, -u_z) ~~~ {\rm and} ~~~ (\omega_x, \omega_y, \omega_z) \rightarrow (\omega_x, -\omega_y, -\omega_z)
\end{equation}

\section{Results}
\label{sec:results}

\subsection{Overview of reconnection}
\label{subsec:overview}

%%%%%%%%%%%%%%%%%%%%%%%%%%%%%%%%%%%%%%%%%%%%%%%%%%%%%%%%%%%%%%%
%% Figure 2 Vortex Topology Changes

\begin{figure}[tb]
\begin{center}
\subfigure[]{
\includegraphics[width=7.0cm]{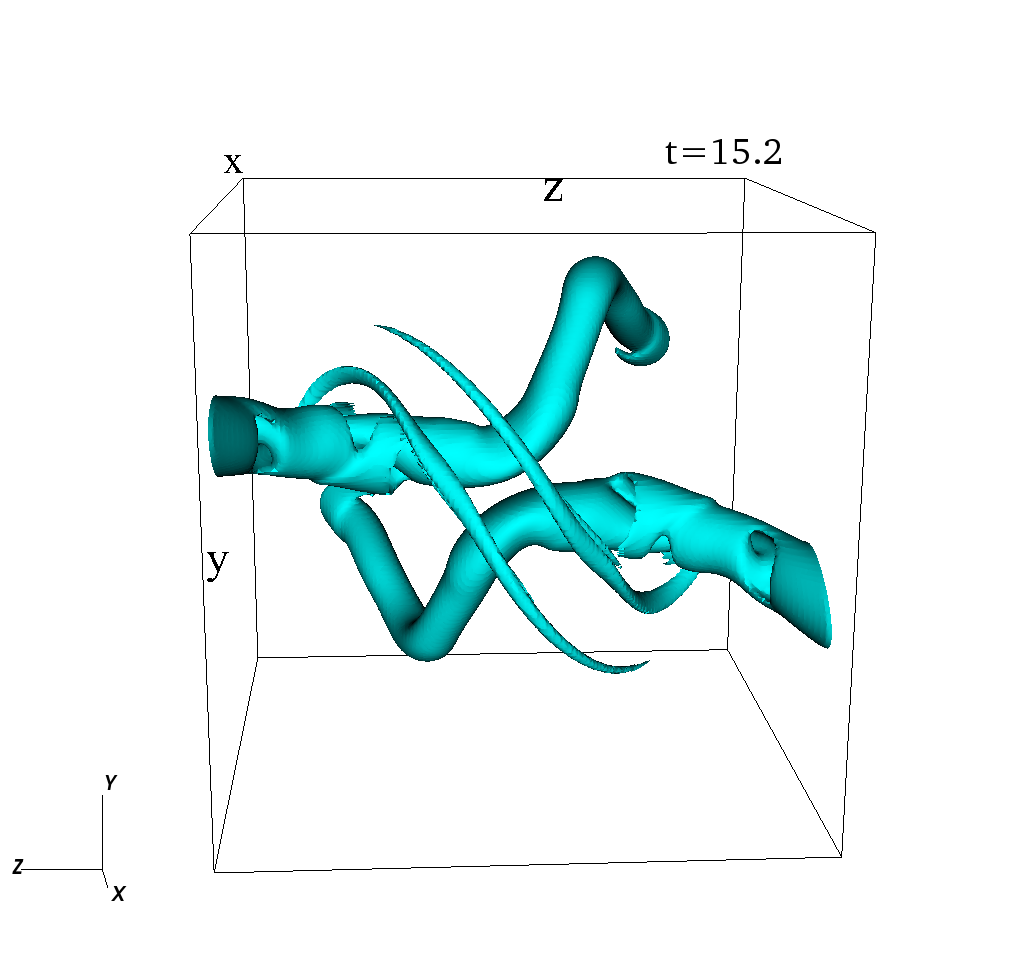}
}
\subfigure[]{
\includegraphics[width=7.0cm]{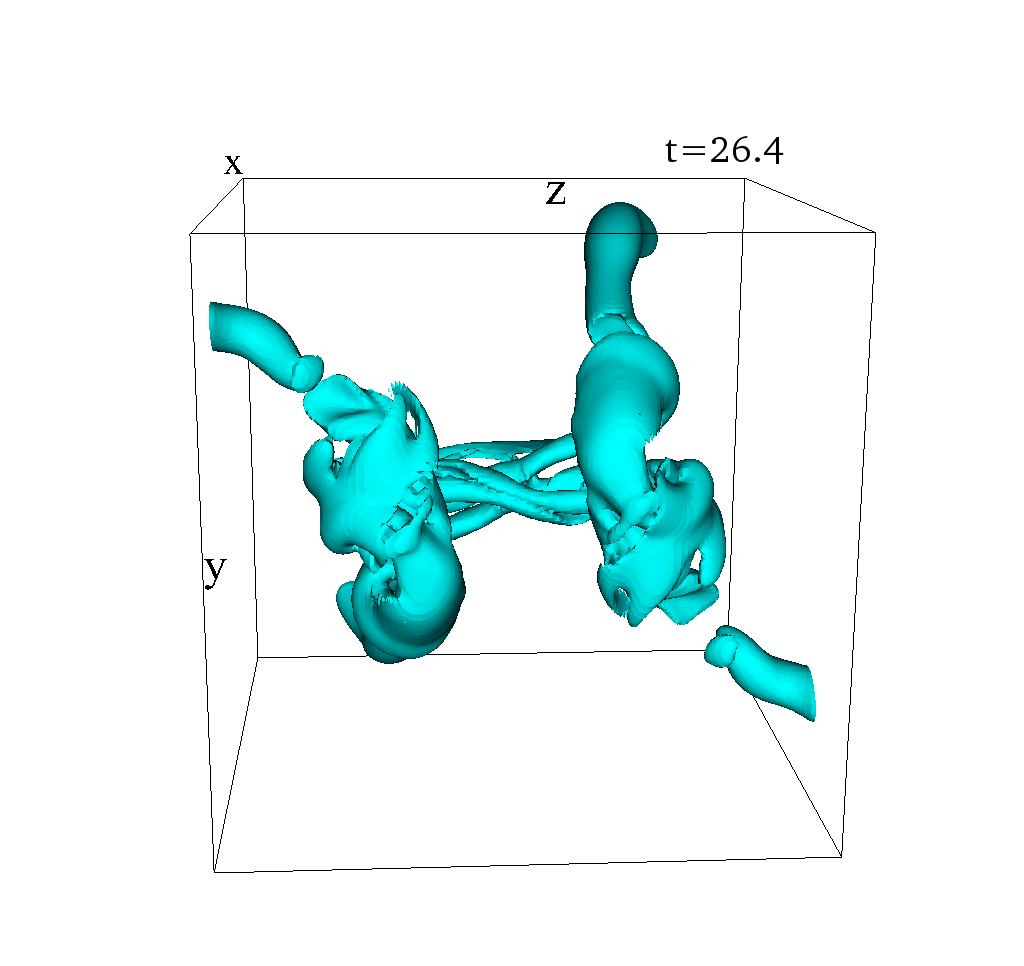}
}
\subfigure[]{
\includegraphics[width=7.0cm]{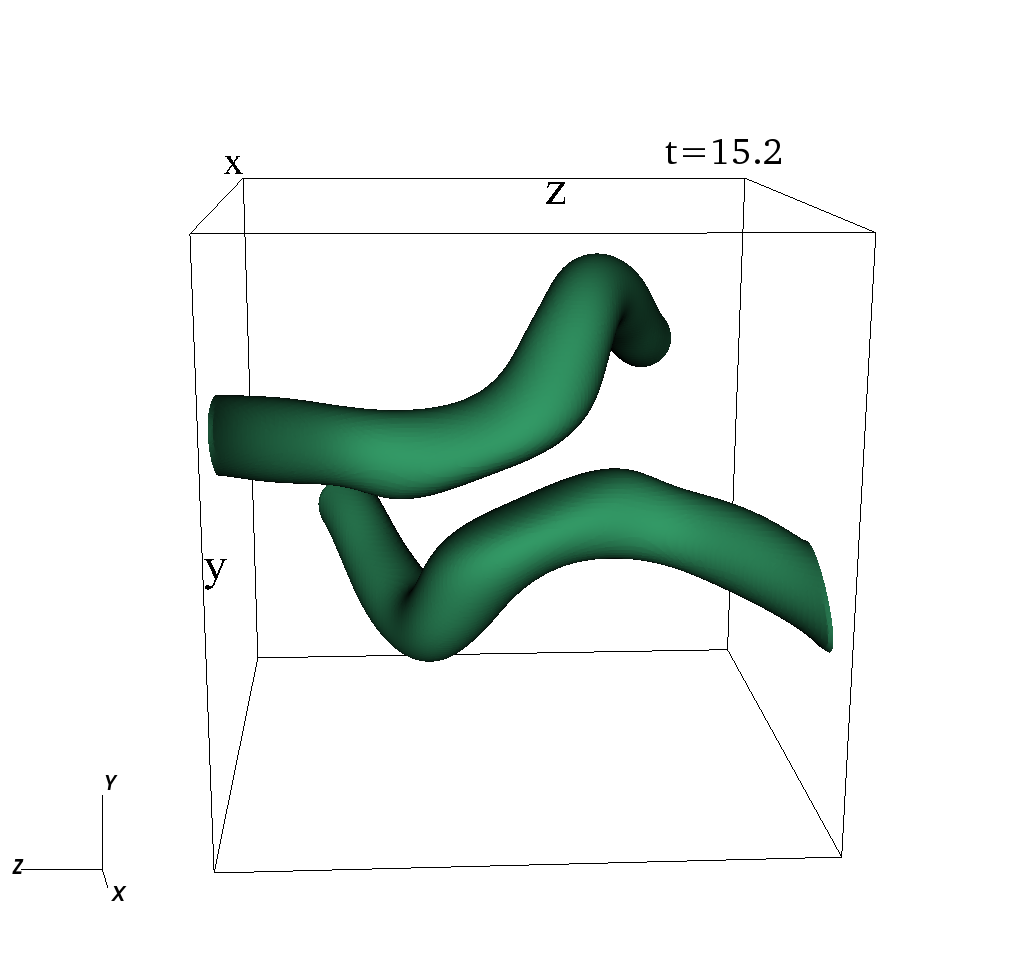}
}
\subfigure[]{
\includegraphics[width=7.0cm]{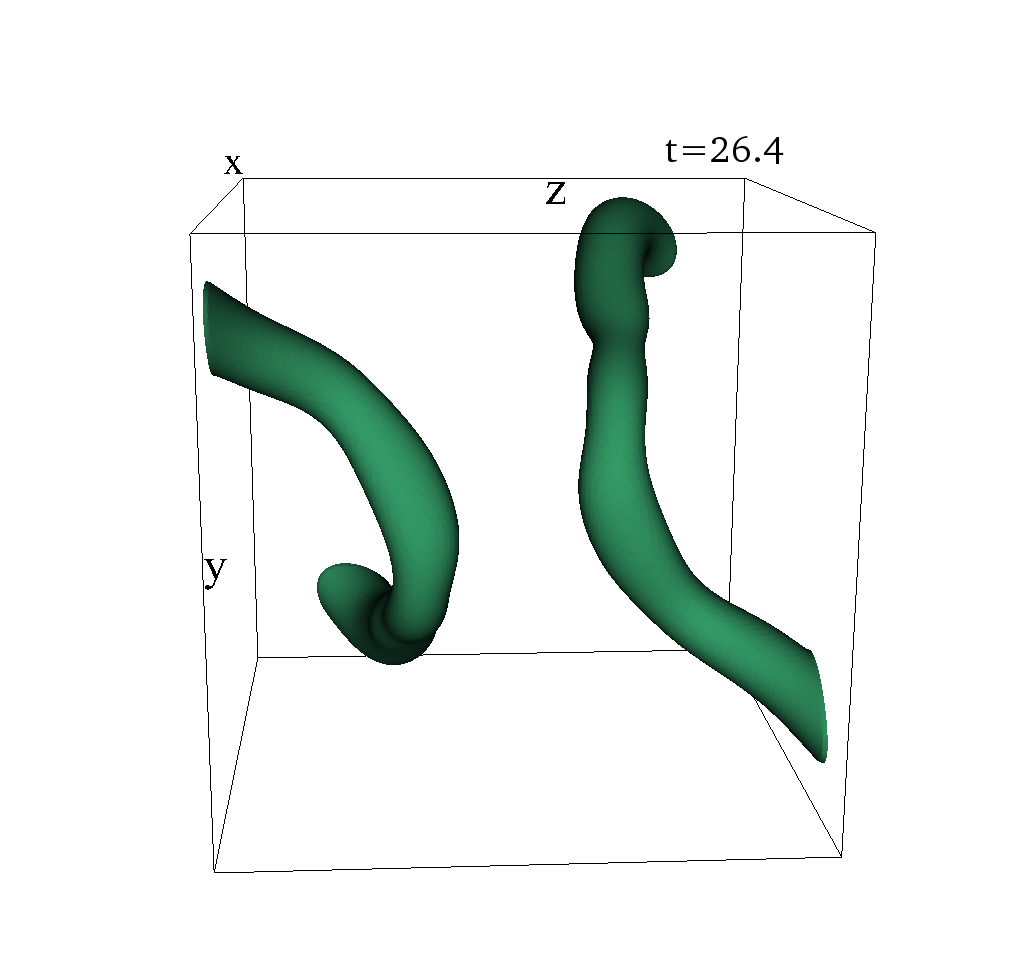}
}
\caption{
Vortex topology changes during a reconnection between initially perpendicular vortex tubes
($b = 1$ and $Re_\Gamma = 4000$, run 3 in  Table~\ref{tab:1}). 
Panels (a) and (b) show iso-vorticity contours corresponding
to $\omega_{th} = 2.5$ at $t = 15.2$, when the tubes begin to interact, and at $t = 26.4$,
after the interaction. These dynamics reveal a change of topology of
the vortex lines and the generation of small-scale vortices during the reconnection.
Panels (c) and (d) show the same vorticity field, band-pass
filtered between $k_< = \sqrt{2} k_f$ and $k_> = 2 \sqrt{2} k_f$, with $k_f = 2.3$, which removes the small-scale features of the flow and showcases only the change of topology of the vortex tubes. The value of the isosurface is $1$ in panels (c) and (d). Full videos of the process are available as supplementary material~\cite{SM}.
}
\label{fig:recon}
\end{center}
\end{figure}
%%%%%%%%%%%%%%%%%%%%%%%%%%%%%%%%%%%%%%%%%%%%%%%%%%%%%%%%%%%%%%%

We begin by illustrating the phenomenon of reconnection for initially perpendicular vortex tubes by
analyzing run 3, with $\beta = 90^\circ$ ($b=1$), $Re_\Gamma = 4000$.
The vortices evolve in time from the initial horizontal conditions, 
as they approach one another and begin to deform. 
Fig.~\ref{fig:recon}(a) shows  the vorticity isosurface at $t=15.2$, before the reconnection event.
As the flow evolves further, the vortices reconnect. This 
results in a changed topology, which can be inferred from the structure of the vorticity field shown in panel (b) at $t=26.4$.
Vertical vortex structures, however, appear simultaneously with smaller scale, horizontal filaments perpendicular to the main tubes. To better characterize the large scale flow structures, present
before and after reconnection, we use the methods applied by Goto et al.~\cite{Goto:2017}, which consists of band-pass filtering the 
vorticity field. For the purpose of the present work, we found it convenient to
isolate the wavenumbers in the band defined by
$ \sqrt{2} k_f \le k \le 2 \sqrt{2} k_f$, where $k_f = 2.3$. This 
filters out the small-scale features clearly seen in panel (a,b),
but leaves apparent the change of topology due to the evolution: the originally
horizontal tubes, parallel to the $z$ axis in 
Fig.\ref{fig:recon}(c), become parallel to the $y$ axis
at later times (see Fig.~\ref{fig:recon}(d)), signaling a topological transition in the vorticity field. 
Between the two times shown in panels (a) and (b), we can observe not only changes in the large-scale vortex topology, but also the progressive appearance of small-scale vortical flow structures. 
At early times, we can observe the appearance of slender vortex filaments which are perpendicular
to the primary vortex tubes and contain little circulation. 
These filaments are a well-known feature of reconnection which has already been discussed
in previous studies, usually under the name of bridges~\cite{Kida:1987,Melander:1988,Kida:1994}. 
In our simulation, they are clearly visible at $t = 15.2$, Fig.~\ref{fig:recon}(a). 
While the vorticity in the bridges is very much amplified at early times through vortex stretching, the role that these slender filaments have in the subsequent interaction of the main tubes during reconnection appears to be limited because they contain little circulation.
Conversely, the small scale features clearly visible at later
times ($t = 26.4$, Fig.~\ref{fig:recon}(b) appear
to be a reproducible feature of the interaction at high Reynolds numbers; 
recent DNS of a pair of vortex tubes, with imposed symmetry
with respect to planes $P_1$ and $P_2$ (in our terminology), also led to 
the formation of a similar small scale structures, which was interpreted to be the 
result of a cascade~\cite{Yao:2020a}. 

\begin{figure}[tb]
\begin{center}
\subfigure[]{
\includegraphics[width=7.0cm]{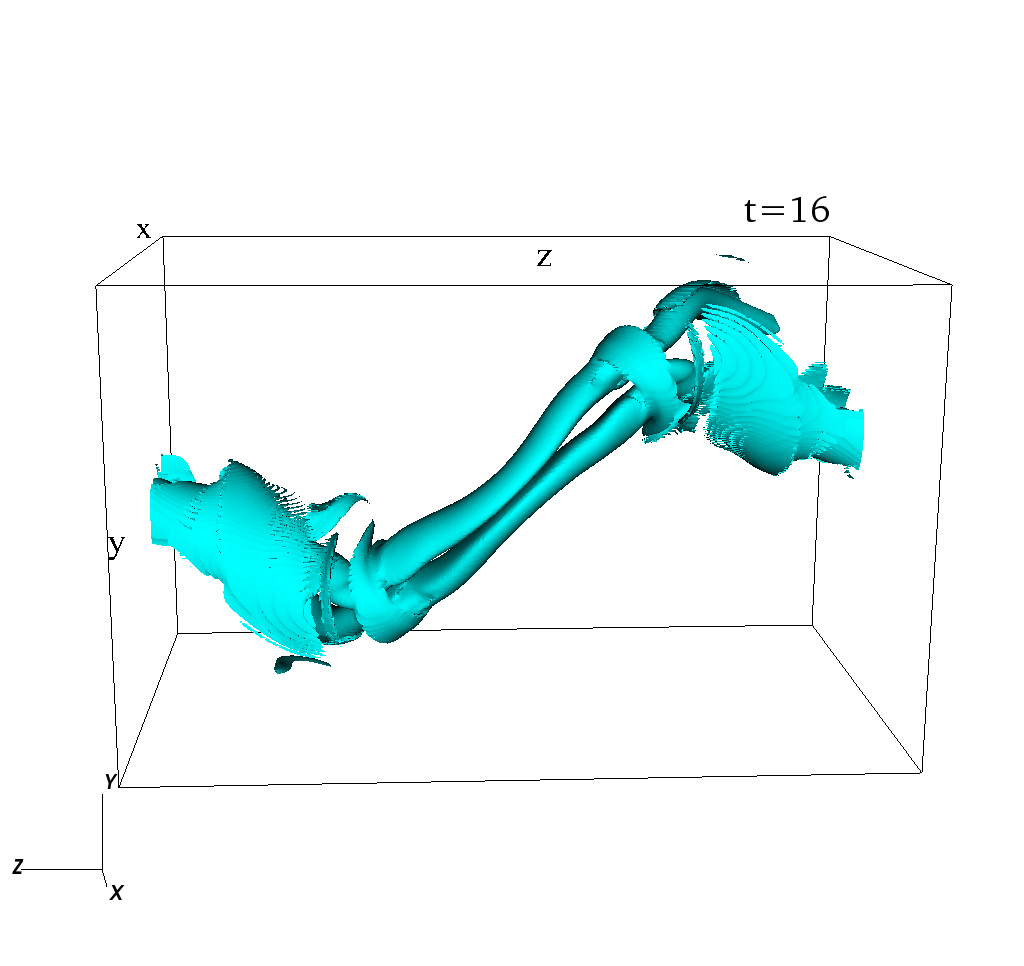}
}
\subfigure[]{
\includegraphics[width=7.0cm]{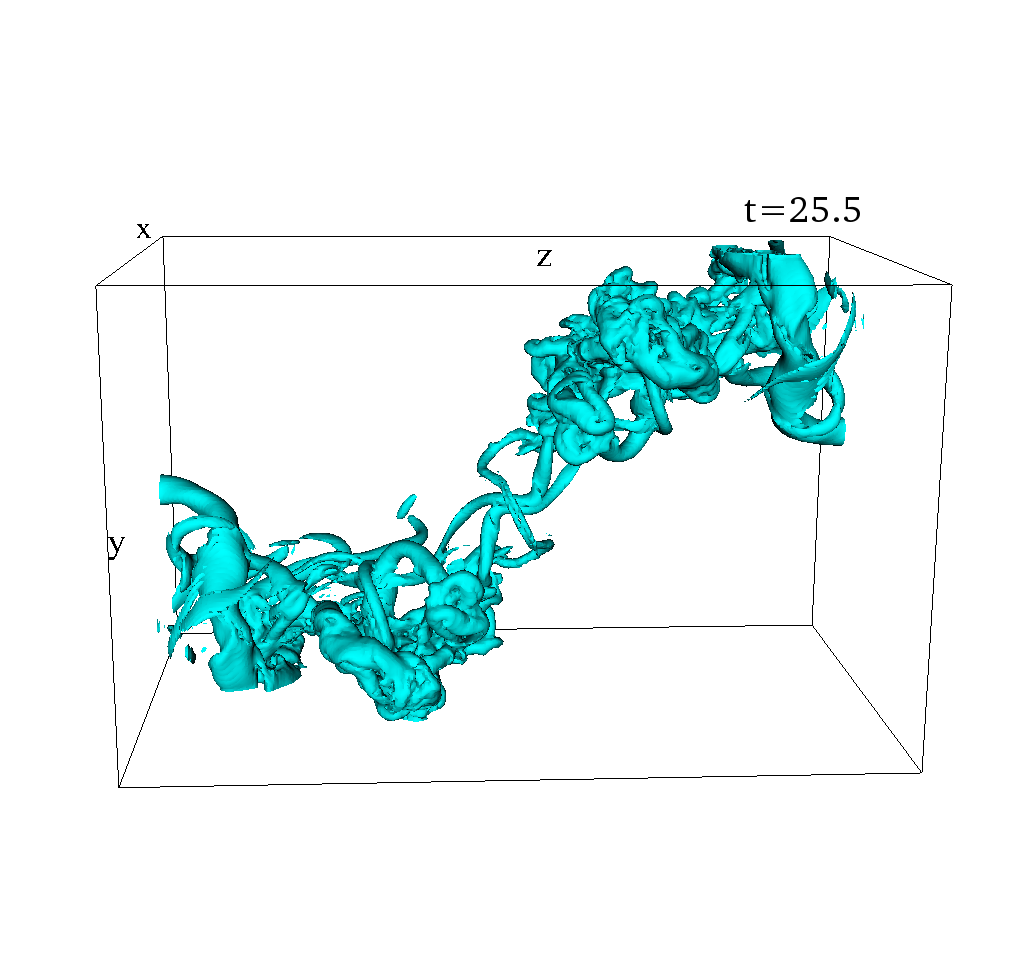}
}
\subfigure[]{
\includegraphics[width=7.0cm]{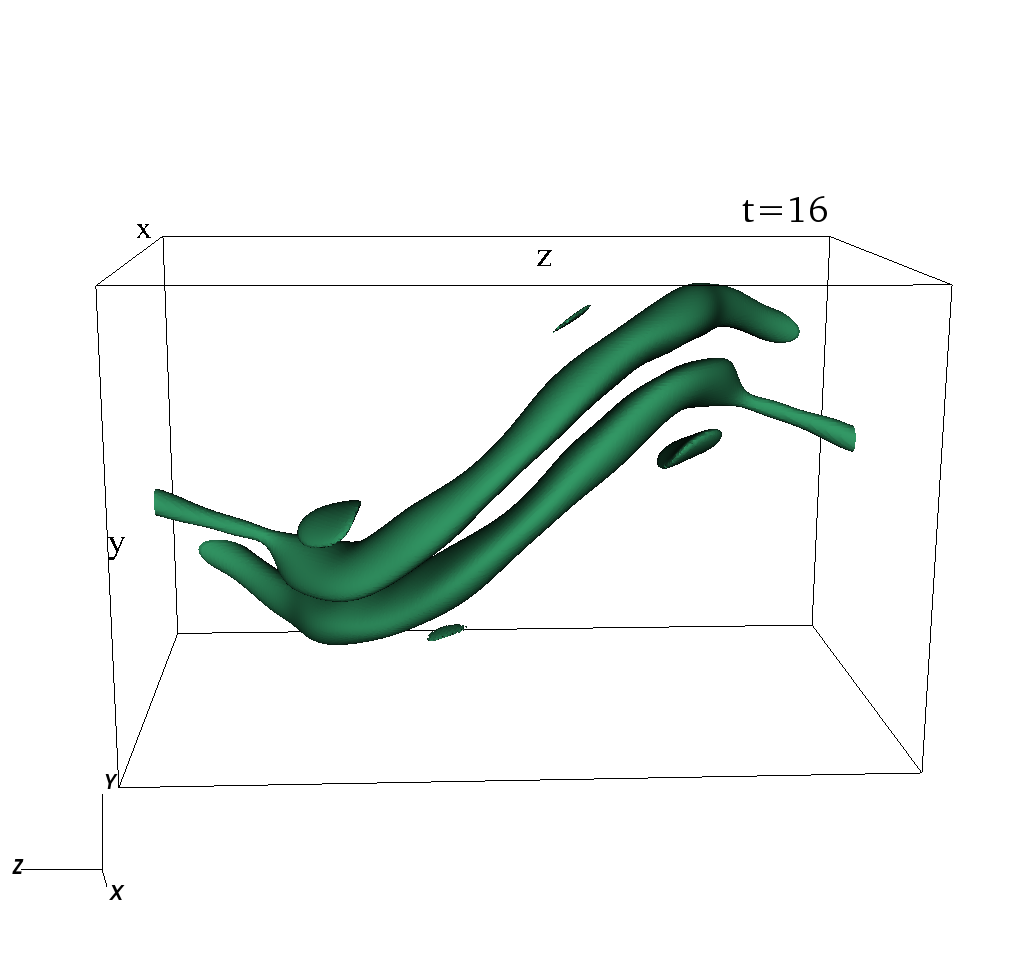}
}
\subfigure[]{
\includegraphics[width=7.0cm]{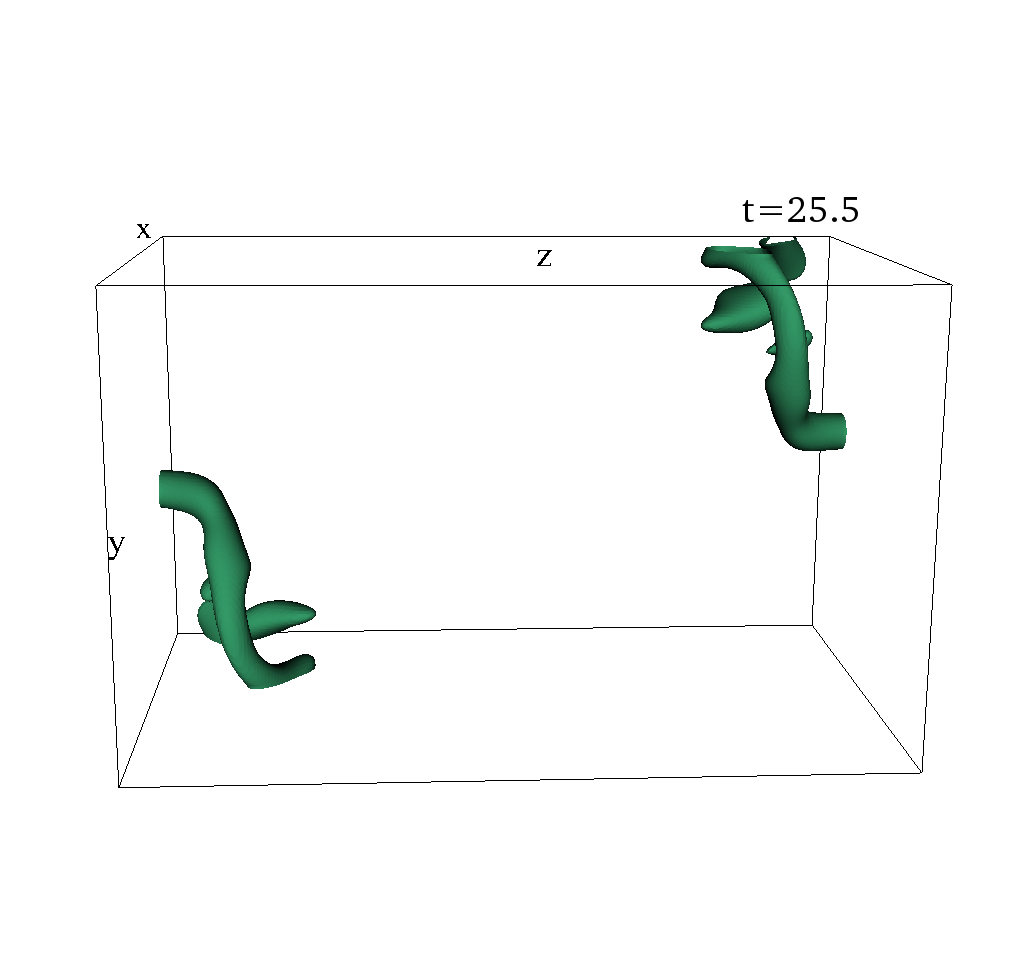}
}
\caption{
Vortex topology changes during a reconnection between tubes at an acute angle
($\beta \approx 28.1^\circ$, $b = 4$ and $Re_\Gamma = 4000$, run 11 in  Table~\ref{tab:1}). 
Panels (a) and (b) show iso-vorticity contours corresponding
to $\omega_{th} = 2.35$ at $t = 16$, when the tubes are paired, and at $t = 25.5$, after the interaction, respectively. These panels show the generation of fine-scale vortices, similar to the collision of two antiparallel vortices ~\cite{McKeown:2020}.
Panels (c) and (d) show the band-passed vorticity field, 
between the wavenumbers $k_<  = \sqrt{2} k_f$ and 
$k_> = 2 \sqrt{2} k_f$ ( where $k_f = 2.3$), at the same times as (a) and (b). The large-scale portions of the vortex tubes at the middle of the interaction zone break down by $t = 25.5$, and the tubes reconnect at the edges of the interaction zone. The value of the isosurface in panels (c) and (d) is $0.85$. Full videos of the process are available as supplementary material~\cite{SM}.
}
\label{fig:recon_b4}
\end{center}
\end{figure}

The reconnection process between two vortex tubes initially at an angle
$\beta = 90^\circ$, illustrated in Fig.~\ref{fig:recon}, exhibits
similarities with reconnection of two tubes with symmetric initial
conditions~\cite{Pumir:1987,Yao:2020a}, as noted e.g. in~\cite{Boratav:1992},
and explained in more detail below. The dynamics leading to reconnection, 
however, are \textit{not} universal. In fact, 
Fig.~\ref{fig:recon_b4} shows an overview of the interaction between
two tubes initially at a much shallower angle, $\beta \approx 28.1^\circ $ 
($b= 4$, run 11). Fig.~\ref{fig:recon_b4}(a) shows that the interaction occurs
over two closely paired sections of the vortex tubes. 
This is clearly illustrated by Figs.~\ref{fig:recon_b4}(c) and (d), which show
iso-vorticity contours of the band-pass filtered solution for $\sqrt{2} k_F \le k \le 2 \sqrt{2} k_F$, with $k_F = 2.3$, as shown in Figs.~\ref{fig:recon}(c) and (d). The configuration with $\beta = 28.1^\circ$ ($b=4$) exhibits a different type of dynamics, resulting in a larger portion of the vortices coming closer to each other than at $\beta = 90^\circ$ ($b = 1$). 
The collision between
these vortex tubes leaves behind a tangle of smaller vortices,
reminiscent of the breakdown that results from the collision of two 
vortex rings~\cite{McKeown:2018,McKeown:2020}. 
This points to a dependence on the initial orientation angle, which is examined further in the following sections.

%%%%%%%%%%%%%%%%%%%%%%%%%%%%%%%%%%%%%%%%%%%%%%%%%%%%%%%%%%%%%%%
%% Figure 3 Vorticity Isosurfaces for acute angle collision

%%%%%%%%%%%%%%%%%%%%%%%%%%%%%%%%%%%%%%%%%%%%%%%%%%%%%%%%%%%%%%%

\subsection{Evolution of two nearly perpendicular tubes: sheet formation. }
\label{subsec:sheets}

\begin{figure}[tb]
\begin{center}
\subfigure[]{
\includegraphics[width=5.15cm]{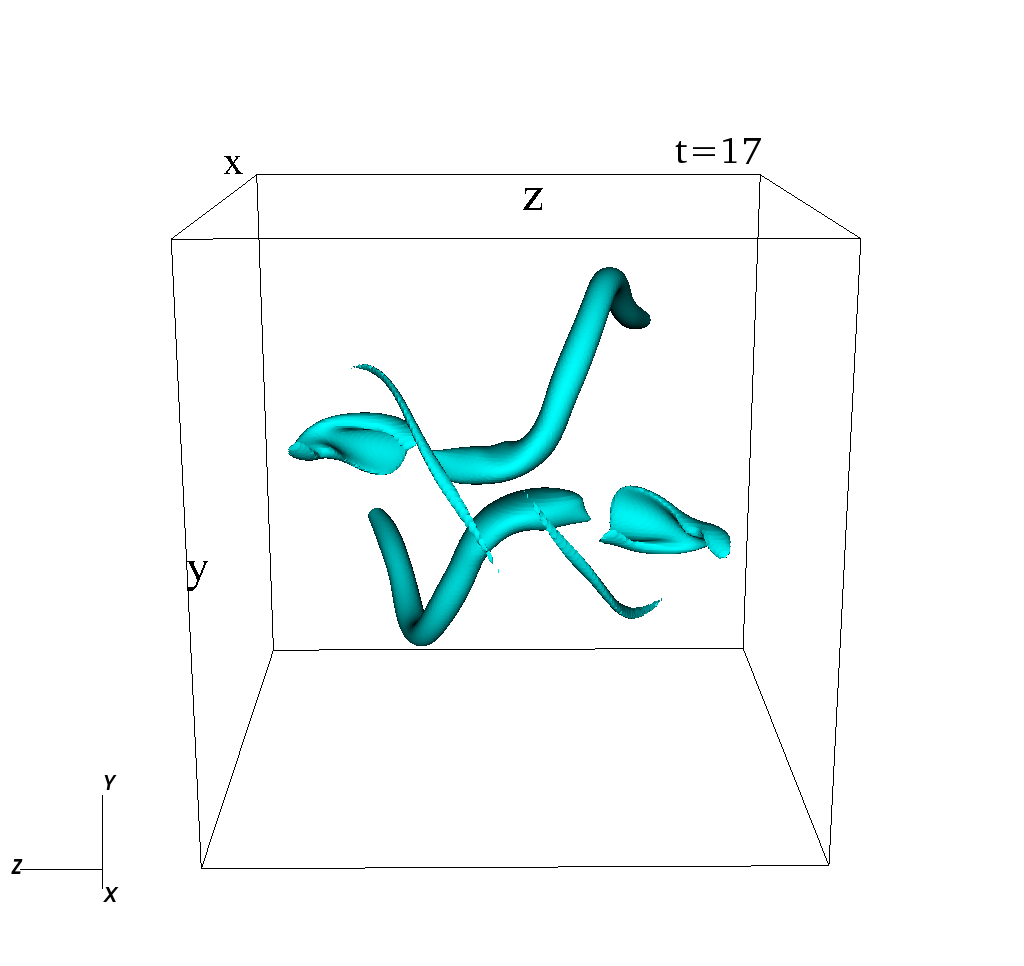}
}
\subfigure[]{
\includegraphics[width=5.15cm]{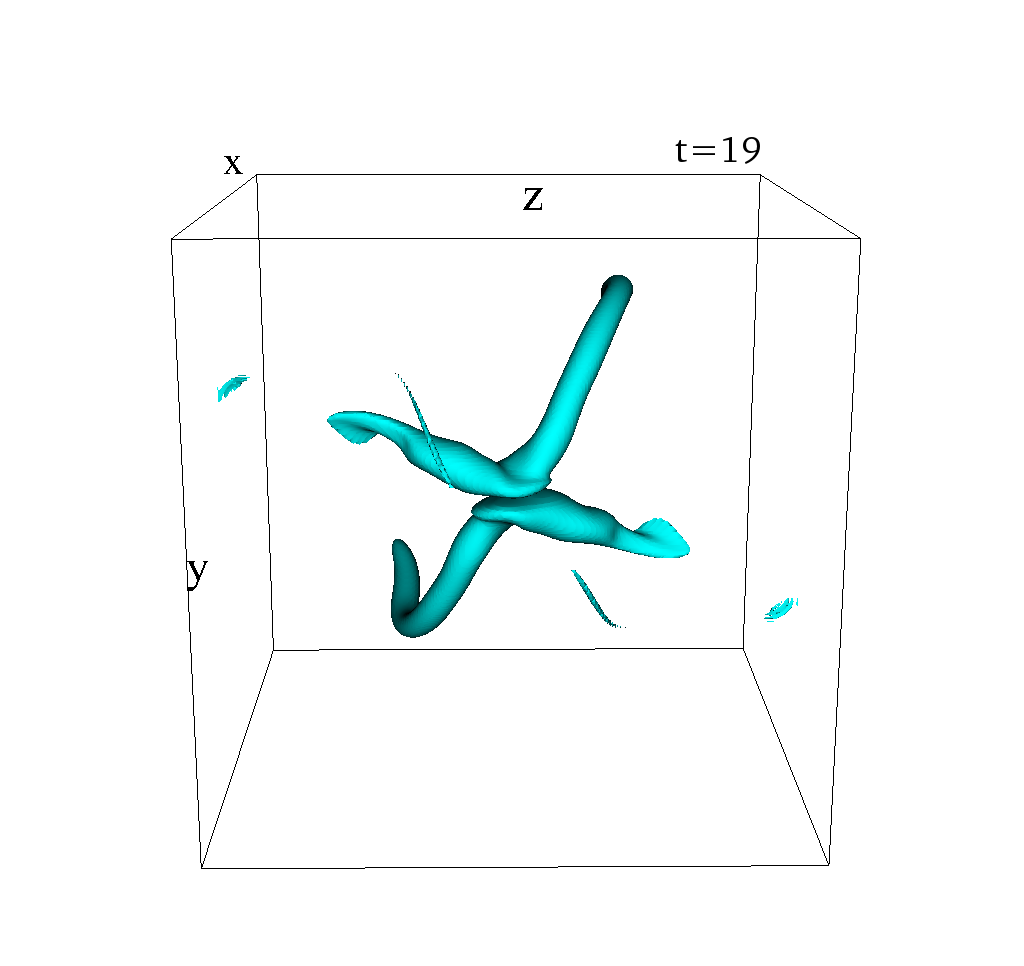}
}
\subfigure[]{
\includegraphics[width=5.15cm]{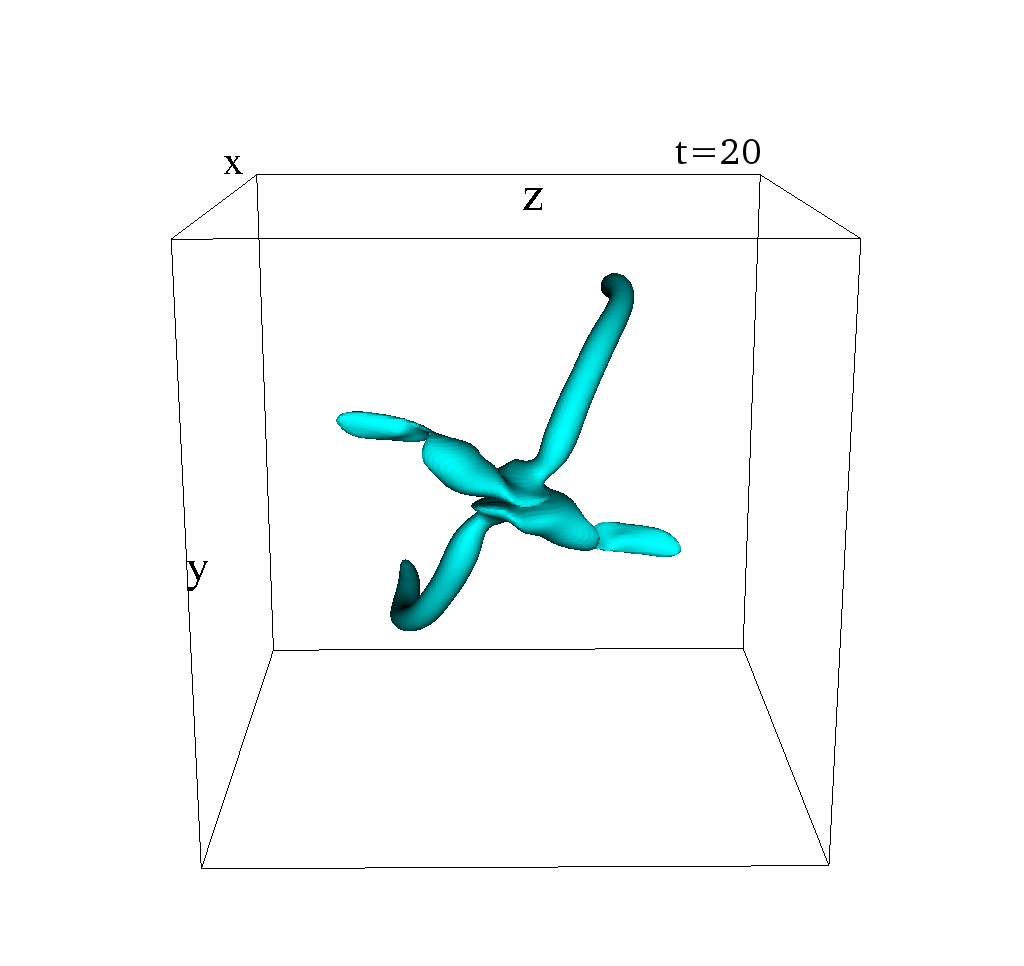}
}
\subfigure[]{
\includegraphics[width=5.15cm]{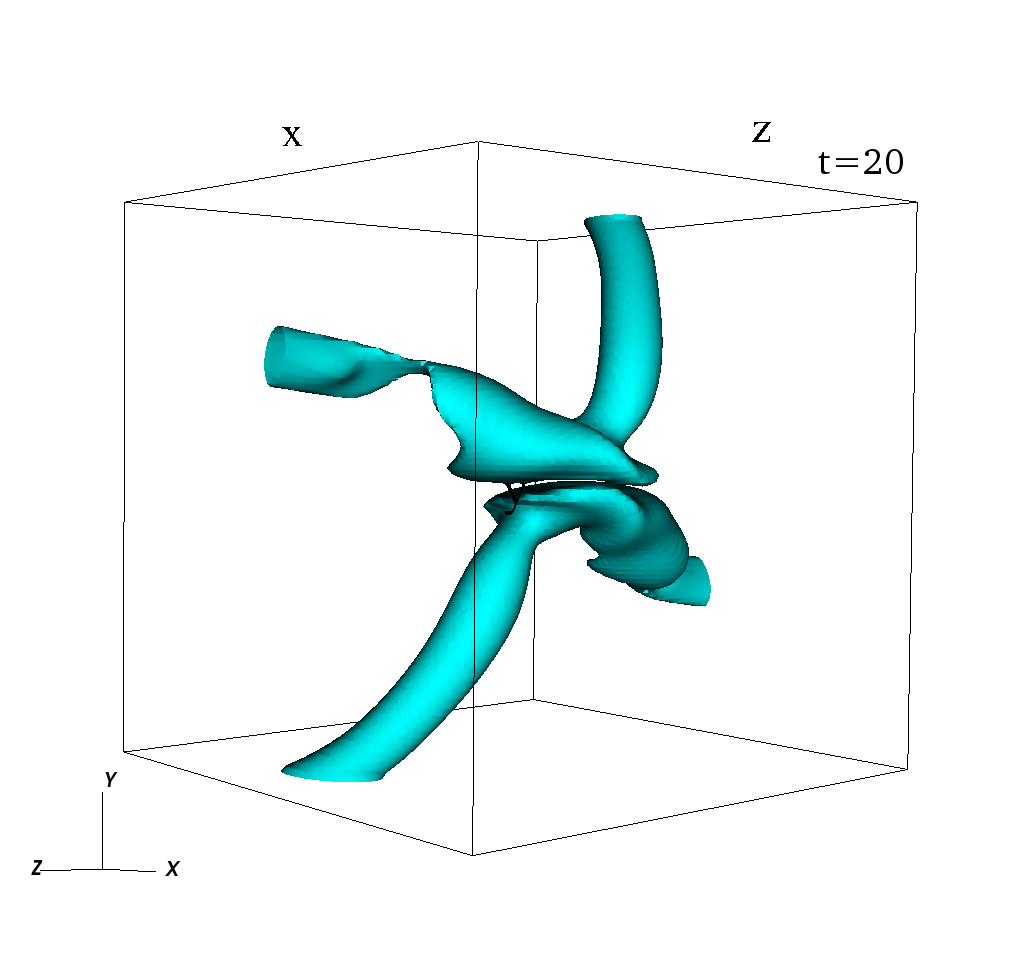}
}
\subfigure[]{
\includegraphics[width=5.15cm]{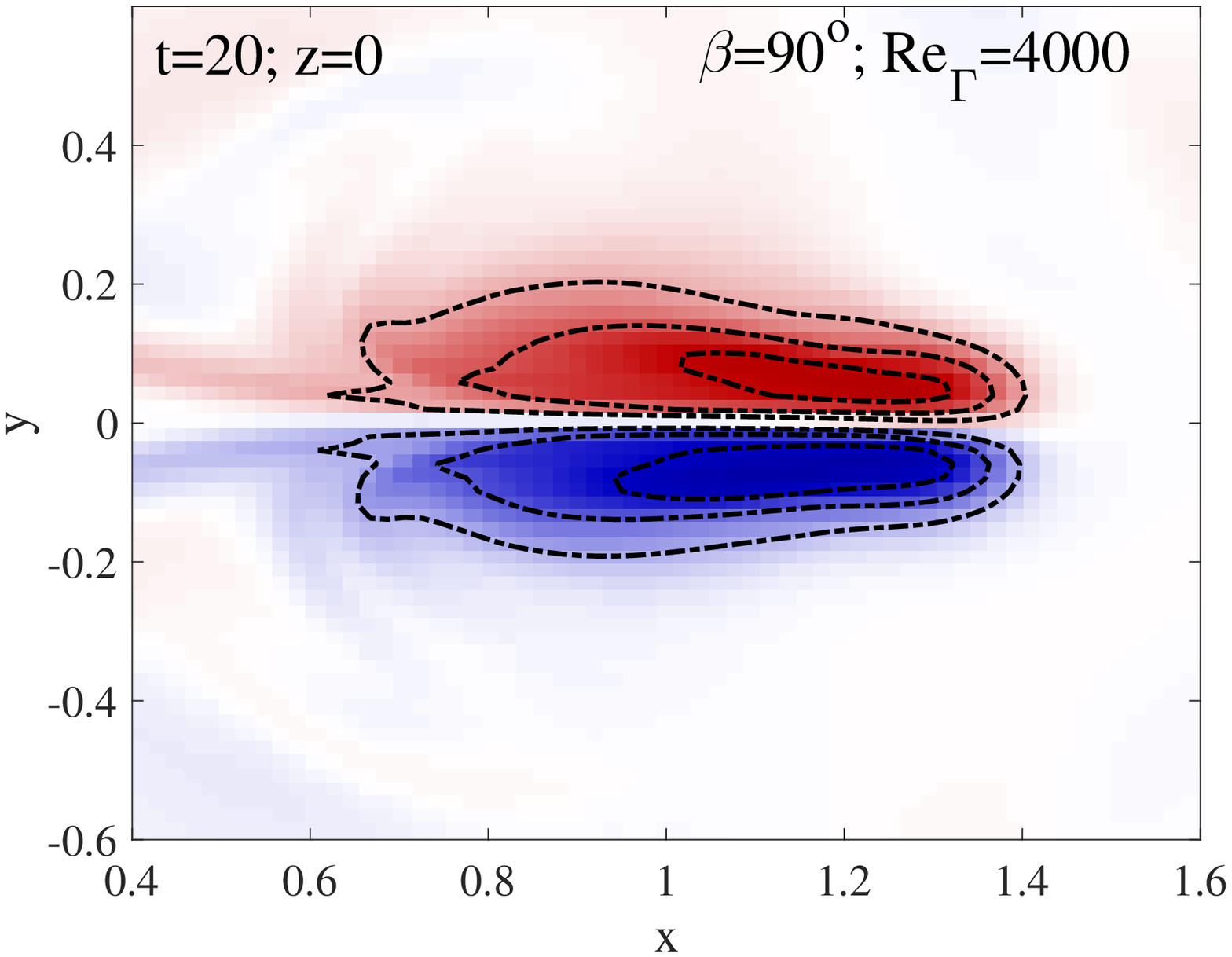}
}
\subfigure[]{
\includegraphics[width=5.21cm]{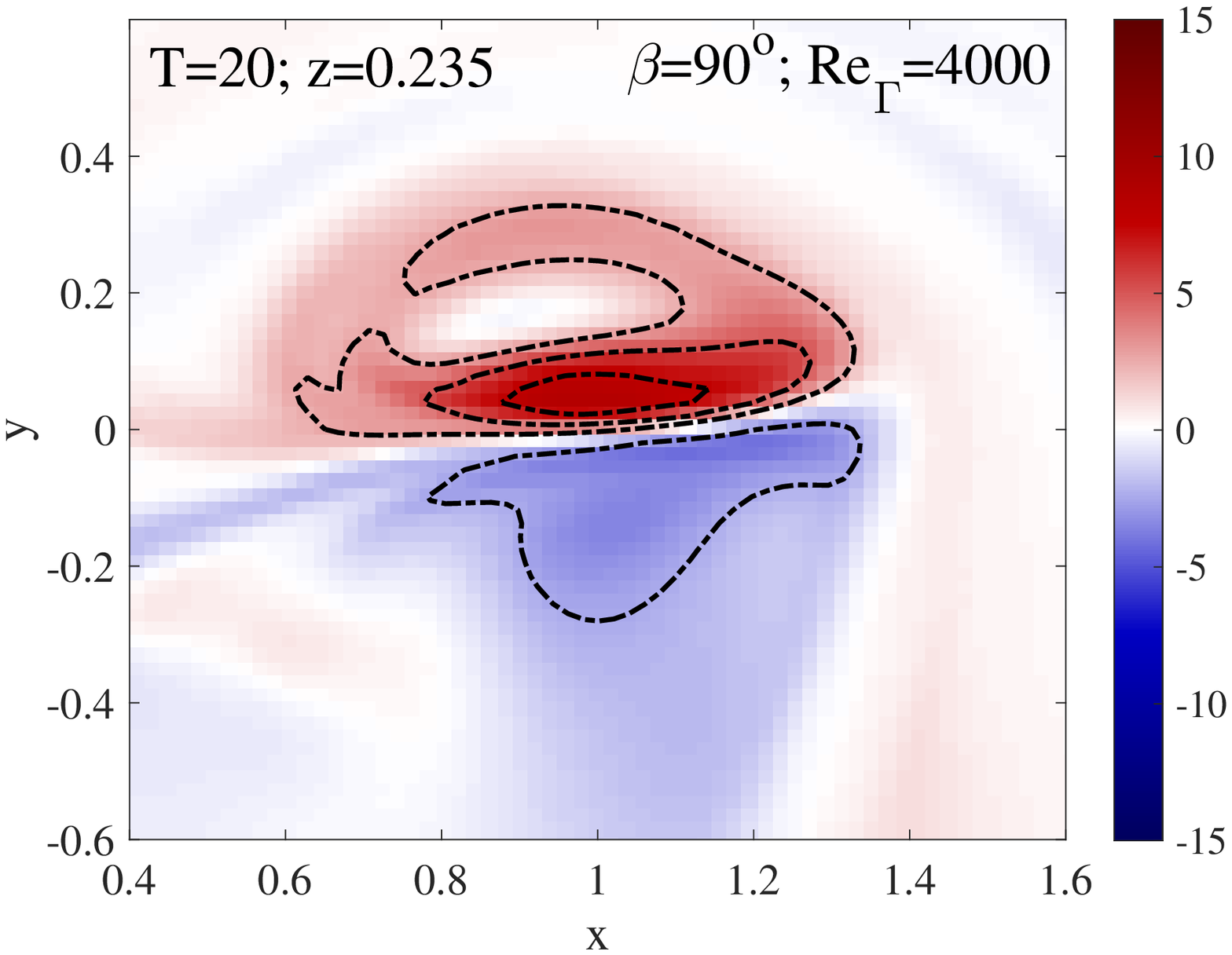}
}
\caption{Close-up view of sheet formation near the reconnection event of initially perpendicular vortex tubes (run 3, $\beta = 90^\circ$, $b = 1$, $Re_{\Gamma} = 4000$).
Iso-vorticity contours of the flow, shown at $t =  17$ (panel (a), $\omega_{th} = 3.6$), $t = 19$ 
(panel (b), $\omega_{th} = 3.8$) and 
$t = 20$ (panel (c), $\omega_{th} = 4.0$), reveal the formation
of a flattened region of amplified vorticity where the two tubes locally contact.
(d) A more detailed view of the region where the vortices interact is 
shown by zooming in and rotating the domain at $t = 20$, $\omega_{th} = 4.0$. (e-f) Cross-sectional plot of the z-component of vorticity in the middle plane, $P_1$ at $z = 0$ (panel e)
and at $z \approx 0.15$ (panel f).
}
\label{fig:inter_b1}
\end{center}
\end{figure}

\paragraph*{Early stage and sheet formation}
The reconnection process starts with the pairing of the tubes, which
locally aligns the vortices in an antiparallel manner, a feature clearly observed directly from the
Biot-Savart equation~\cite{Siggia:1985}, and consistent with all previous numerical observations. 
The local pairing of antiparallel filaments is accompanied  by a significant deformation of the vortex tubes.
Fig.~\ref{fig:inter_b1}(a-c) shows iso-surfaces
of the vorticity magnitude for $\beta = 90^\circ$ and $Re_\Gamma = 4000$ to illustrate this interaction.
The three views from a perspective similar to that shown in 
Fig.~\ref{fig:scheme}(a), at $t = 17$ (Fig.~\ref{fig:inter_b1}(a)),  
$t = 19$ (Fig.~\ref{fig:inter_b1}(b)) and $t = 20$
(Fig.~\ref{fig:inter_b1}(c)) indicate that the nearest regions of the tubes come together
and flatten into thin vortex sheets. It is important to 
notice that the spatial extent of the vortex sheets, in the direction of the 
vortex tubes, is in fact rather limited.
The vortex sheets are confined in the $z$-direction to a size smaller than that of the initial vortex cores. 
We also stress that the 
sheets do not appear to perfectly align with the $P_2$ ($y = 0$) plane, 
as it happens in the canonical problem of two initially weakly perturbed antiparallel 
vortex tubes, symmetric with respect to the midplane. In fact, the tilt of the sheets increases from $t = 17$ to $t = 20$.

%%%%%%%%%%%%%%%%%%%%%%%%%%%%%%%%%%%%%%%%%%%%%%%%%%%%%%%%%%%%%%%
%% Figure 4 Close-up view of sheet formation during reconnection

%%%%%%%%%%%%%%%%%%%%%%%%%%%%%%%%%%%%%%%%%%%%%%%%%%%%%%%%%%%%%%%

The pairing shown in Fig.~\ref{fig:inter_b1}(a-c) with the formation of 
vortex sheets in the regions where the vortices interact, is qualitatively 
consistent with the simulations of~\cite{Boratav:1992}. 
As already stated, the formation of sheets is a robust feature in many 
simulations of interacting vortex tubes, starting with an initial configuration of almost parallel counter-rotating tubes with a slight perturbation~\cite{Kerr:2013}. 
Further insight on these vortex sheets is provided by Fig.~\ref{fig:inter_b1}(d),
which shows a magnified view at $t = 20$, from a slightly different 
perspective, showcasing the pronounced flattening of the vortex cores.
As shown in Fig.~\ref{fig:inter_b1}(e-f), the isocontours of the $z$ component of vorticity
at the central plane, $z = 0$ (the plane $P_1$, as 
introduced in Fig.~\ref{fig:scheme}(a)) and at an adjacent plane parallel to 
$P_1$ which is slightly off the symmetry plane, further indicate the formation of intense, thin vortex sheets.
Up to the time shown in Fig.~\ref{fig:inter_b1}(d), the flattening of the
sheets is not greatly affected by increasing the Reynolds number. We view this
as evidence that the formation of the narrow vortex sheets is only 
a precursor of reconnection.

We notice that Fig.~\ref{fig:inter_b1}(a-c) also demonstrates that the vortex filament ``bridges'', clearly visible at the earlier stages of the interaction when the tubes are drawn closer together, 
(Fig.~\ref{fig:recon}(a)), are still visible at $t = 17$ 
(Fig.~\ref{fig:inter_b1}(a)). As the vortex tubes begin to flatten into sheets at $t = 19$ (Fig.~\ref{fig:inter_b1}(b)), these bridges become less pronounced and are no longer present at $t = 20$ (Fig.~\ref{fig:inter_b1}(c)). Part of the reason why the bridges
are less visible at later times is the increase in the vorticity threshold, $\omega_{th}$
used at the three different times. The vorticity magnitude increases locally at the reconnection site where the cores become locally flattened into sheets.
In fact, the bridges are concentrated in very narrow regions of space; this 
implies that large velocity gradients are generated, but that 
viscosity acts very strongly to dissipate them. 
For these reasons, as already stated, the bridges do not play 
any appreciable role in the reconnection dynamics and are immaterial for the present discussion.

\begin{figure}[tb]
\begin{center}
\subfigure[]{
\includegraphics[width=7cm]{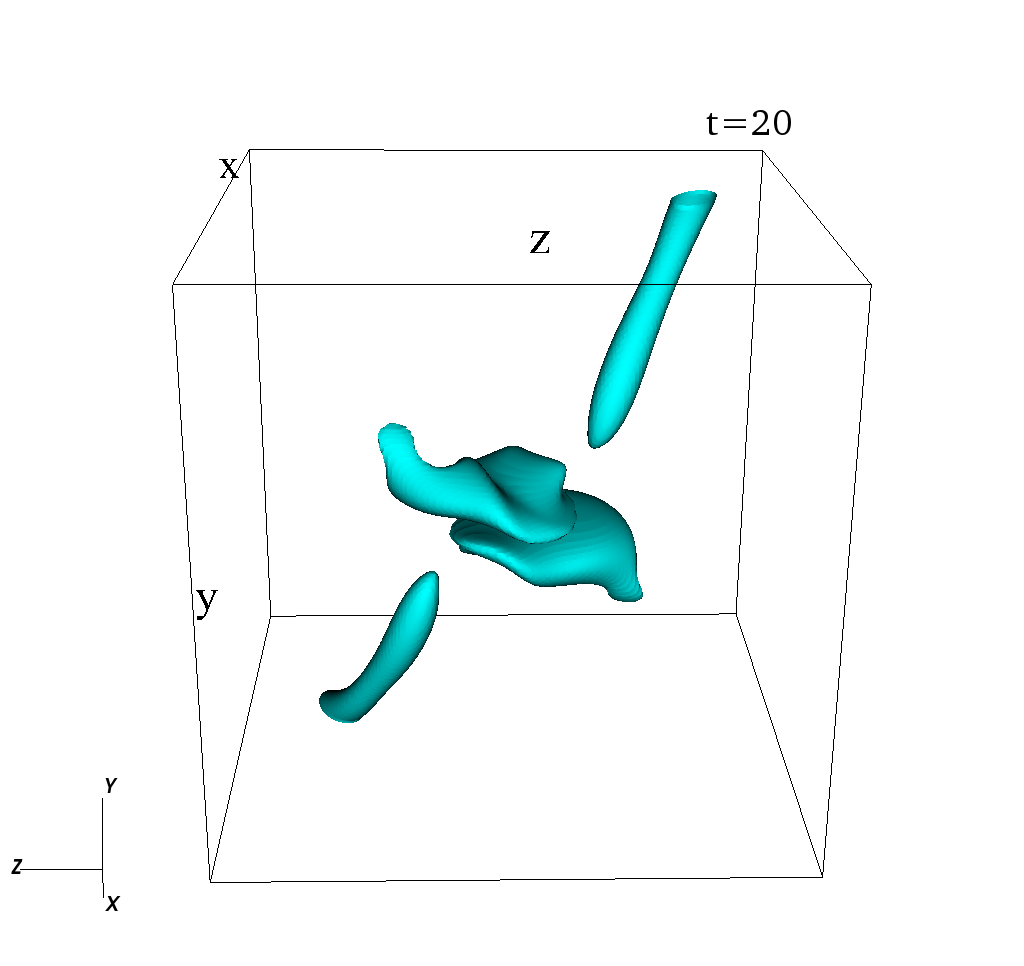}
}
\subfigure[]{
\includegraphics[width=7cm]{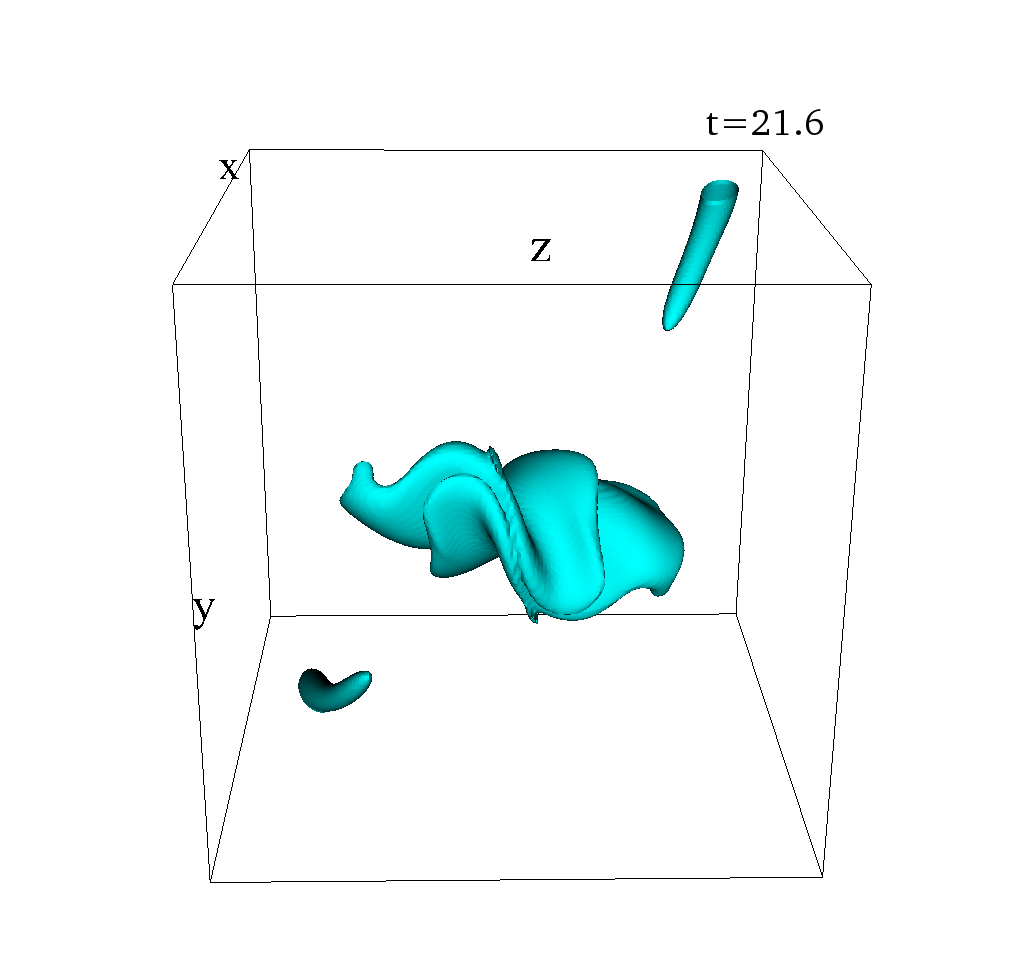}
}
\subfigure[]{
\includegraphics[width=7cm]{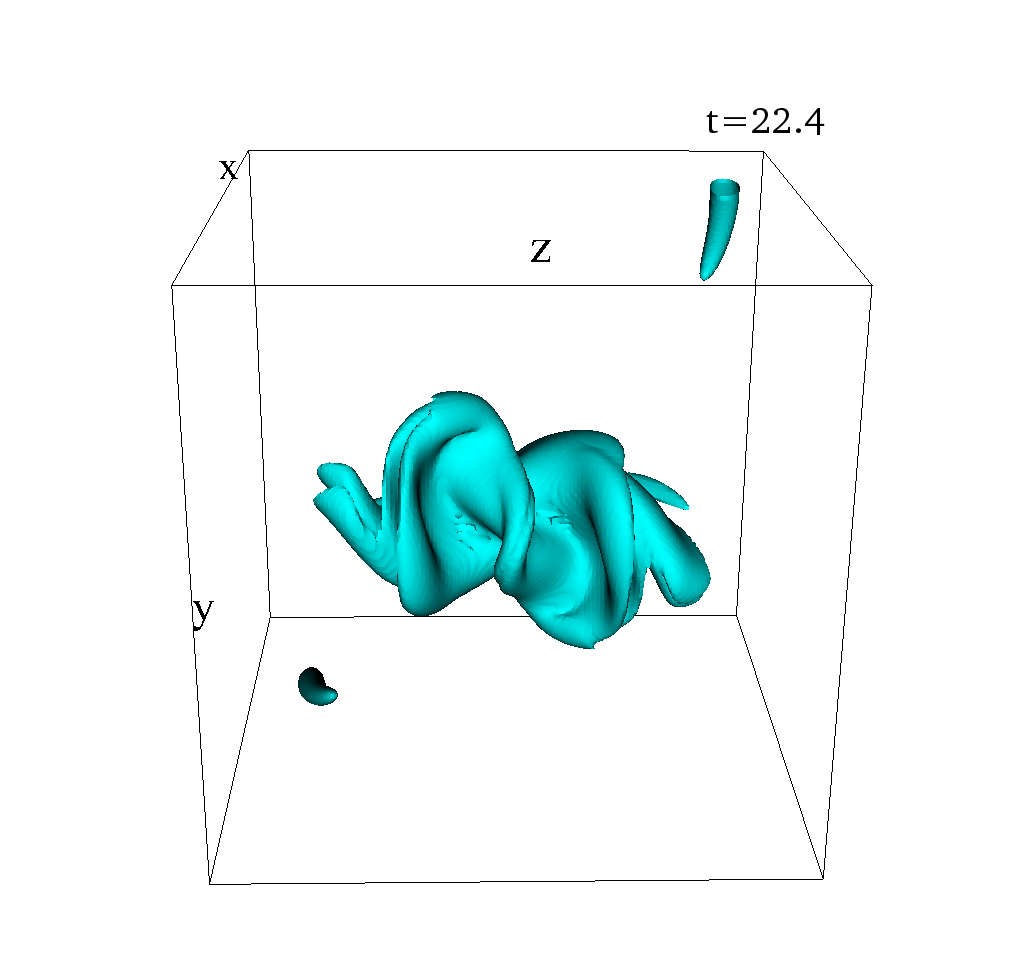}
}
\subfigure[]{
\includegraphics[width=7cm]{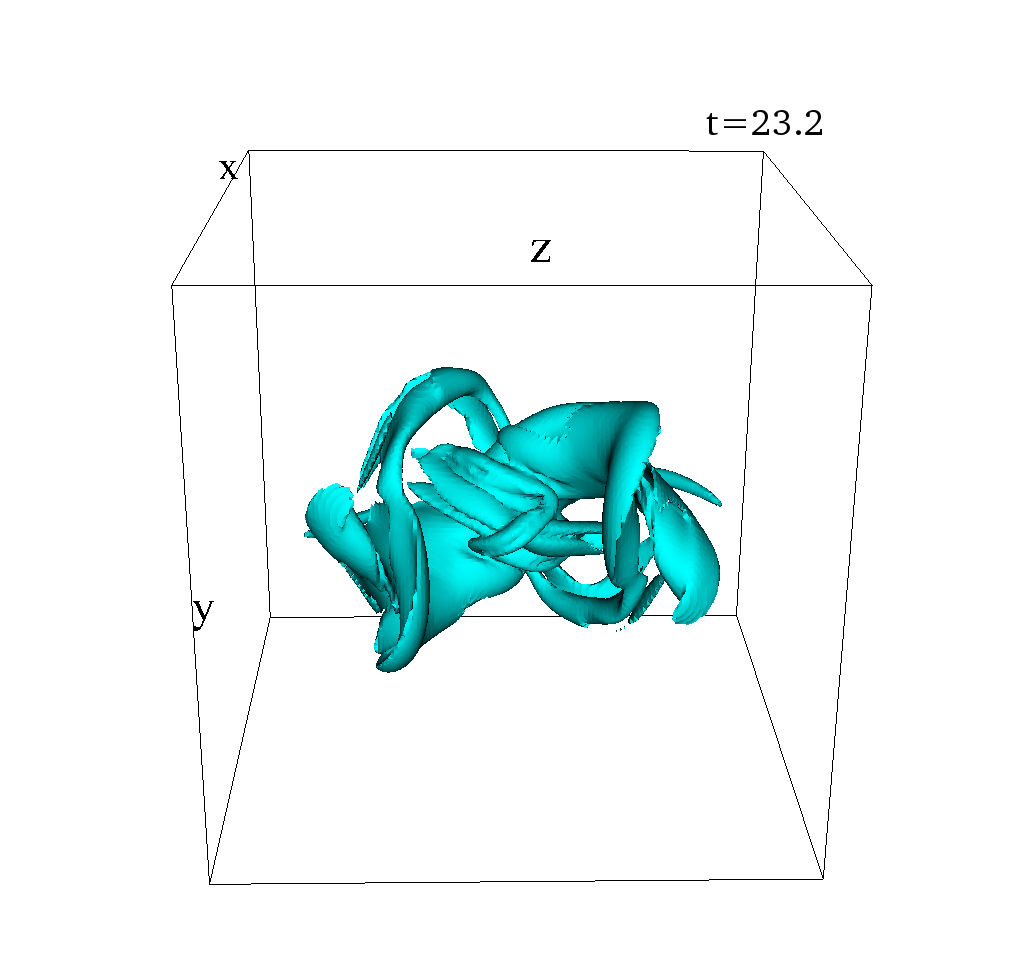}
}
\caption{
Twisting of the vortex sheets and reconnection.
Vorticity magnitude isosurface for initially perpendicular vortex tubes in run 3 ($\beta = 90^\circ$, $Re_{\Gamma} = 4000$)
at three consecutive times, following the vortex sheet formation. 
(a) After the vortex cores flatten into sheets, they become twisted as they wrap around each other. (b) The vortex sheets become folded and reorient along the $z = 0$ ($P_1$) plane, forming an array of transverse vortex filaments. (c) The two strong vortex sheets annihilate, leaving a tangle of small-scale vortex filaments. The last time shown, $t = 23.2$ corresponds approximately to the peak dissipation rate, see Fig.~\ref{fig:Egy_Enst_b1}(b).
The iso-vorticity contours of the flow, are shown at 
$t =  21.6$ ($\omega_{th} = 5.25$), $t = 22.4$ 
($\omega_{th} = 5.60$) and 
$t = 23.2$ ($\omega_{th} = 6.0$). The plane $P_1$ ($z = 0$) separates the box shown in the middle.
}
\label{fig:end_recon_b1}
\end{center}
\end{figure}

\paragraph*{Late stage and reconnection }
The flattening of the cores into  sheets is the precursor of the reconnection process.
Up until the latest time, shown in Fig.~\ref{fig:inter_b1}(c), the vortex 
lines are not broken; they are brought together and compressed into a narrow region. The change of topology of the vortex lines, clearly illustrated in 
Fig.~\ref{fig:recon}, occurs at a later stage through the destruction of the vortex sheets. 
We stress that this process is strongly constrained by the symmetry imposed, as in the previously studied case of the 
two initially weakly perturbed, antiparallel vortex 
tubes (c.f.~\cite{Yao:2020a}). With our initial conditions, as previously noted, the sheets do not particularly align with any plane. In fact,
as shown in Fig.~\ref{fig:end_recon_b1}, the vortex sheets strongly deform 
in a fully 3-dimensional manner at later times. The vortex sheets begin to twist around each other at $t = 21.6$ (Fig.~\ref{fig:end_recon_b1}(a)),
until the sheets, originally aligned mostly parallel to the $P_2$ ($y = 0$) 
plane, become almost vertical along the $P_1$ plane ($z = 0$), as shown in 
Fig.~\ref{fig:end_recon_b1}(b). The continued twisting of the vortex sheets causes them to become locally folded along both sides of the $P_1$ plane, leading to the formation of transverse vortex filaments. Shortly afterward, at $t = 23.2$, 
shown in Fig.~\ref{fig:end_recon_b1}(c), the main sheets in the $P_1$ plane are anihilated, leaving behind a complicated tangle of small-scale vortices.
The results of Fig.~\ref{fig:end_recon_b1} therefore show that once the vortices pair off and begin to interact, the evolution of the reconnection dynamics differs significantly from those
obtained with a much more symmetric initial condition, such as~\cite{Yao:2020a} which cannot account for the twisting of the sheets. This difference may affect the formation of small-scale vortices,
which form at later times, as shown in Fig.~\ref{fig:recon}(b).

%%%%%%%%%%%%%%%%%%%%%%%%%%%%%%%%%%%%%%%%%%%%%%%%%%%%%%%%%%%%%%%
%% Figure 5 Evolution of vortex sheets and reconnection
%%%%%%%%%%%%%%%%%%%%%%%%%%%%%%%%%%%%%%%%%%%%%%%%%%%%%%%%%%%%%%%

%%%%%%%%%%%%%%%%%%%%%%%%%%%%%%%%%%%%%%%%%%%%%%%%%%%%%%%%%%%%%%%
%% Figure 6 kinetic energy, dissipation, and 6th moment of vorticity
\begin{figure}[tb]
\begin{center}
\subfigure[]{
\includegraphics[width=5.17cm]{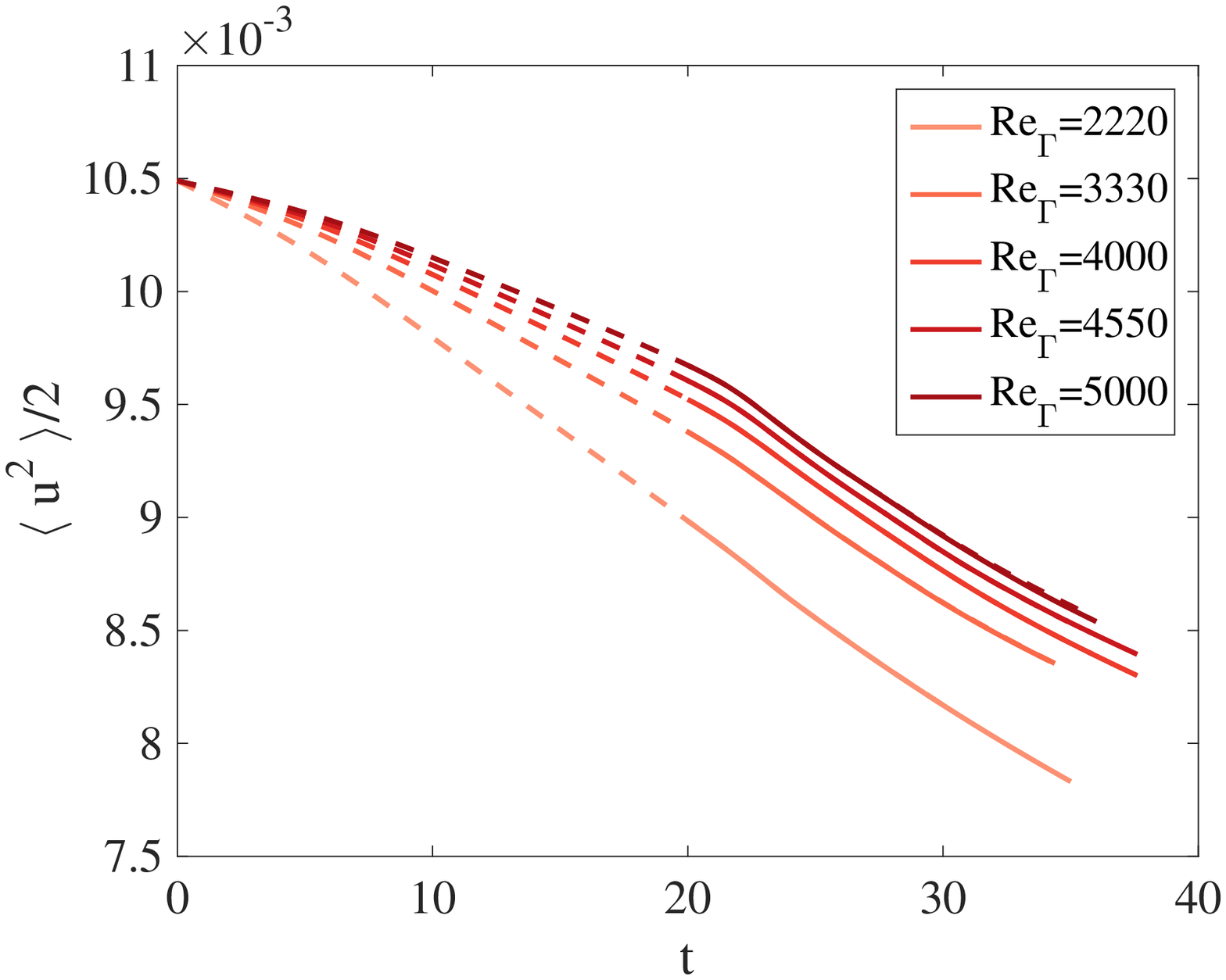}
}
\subfigure[]{
\includegraphics[width=5.17cm]{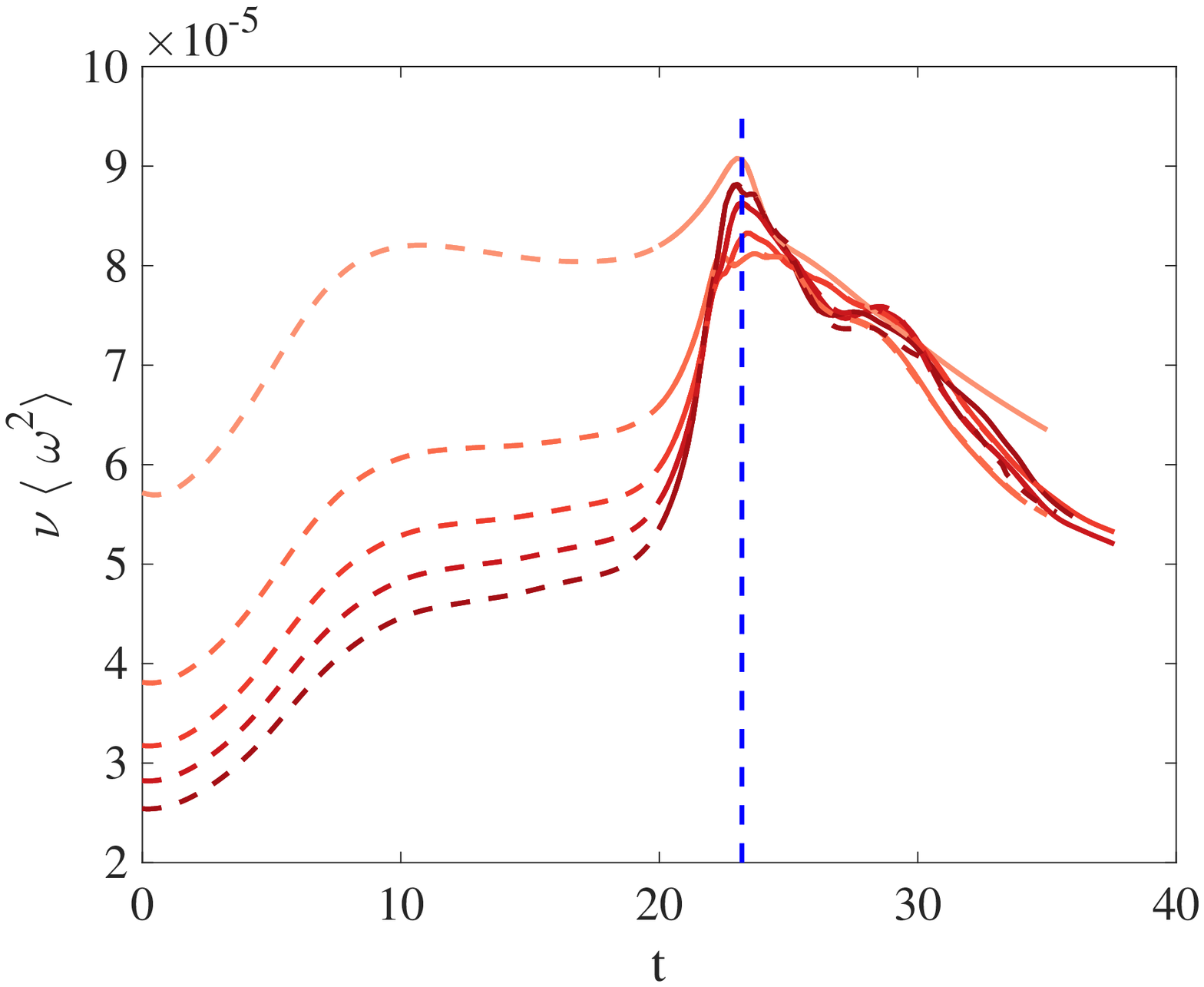}
}
\subfigure[]{
\includegraphics[width=5.17cm]{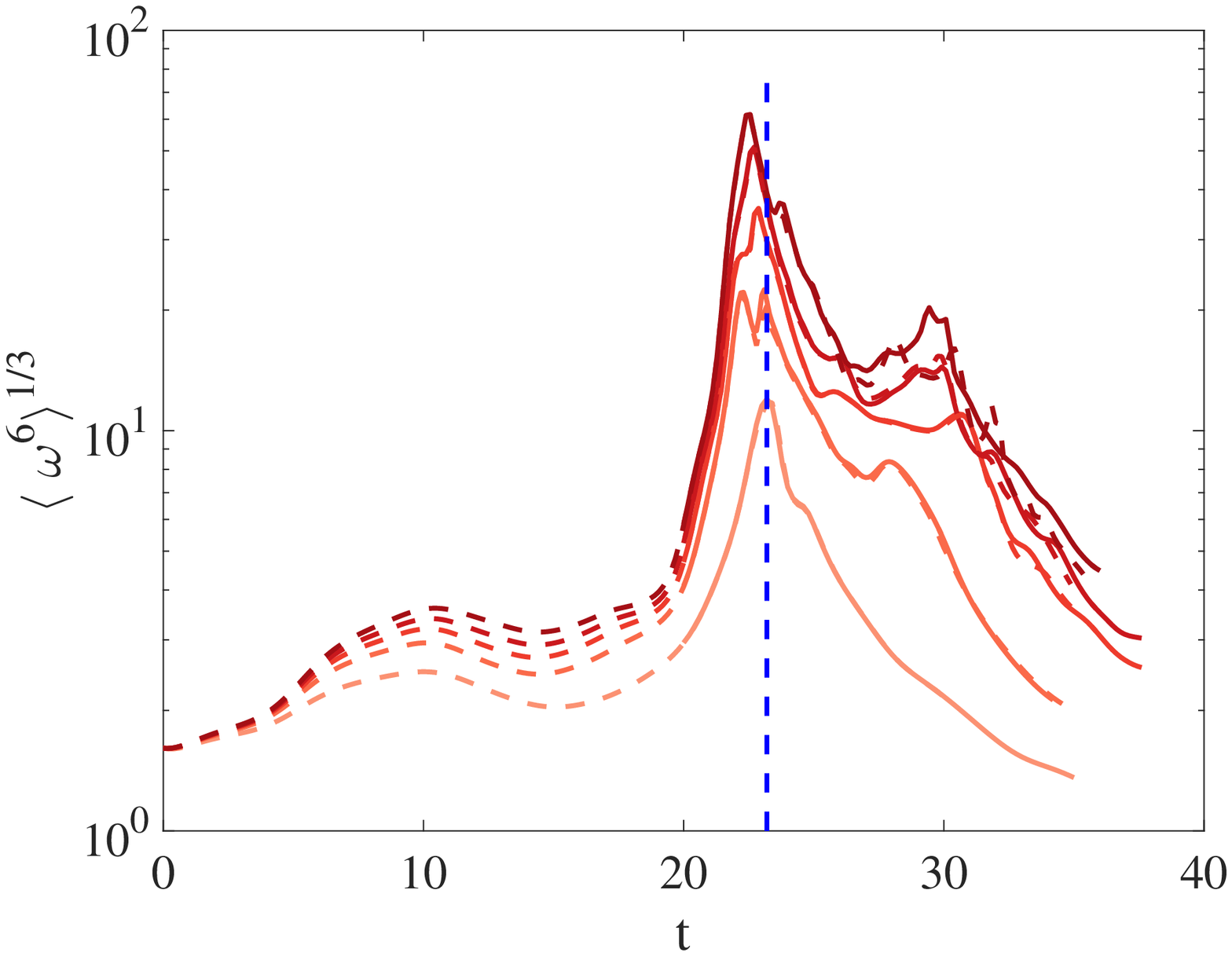}
}
\caption{
Global dynamics for initially perpendicular vortex tubes at various Reynolds numbers. The evolution of (a) the mean kinetic energy rate, $\langle \ve u ^2 \rangle /2$, (b) the mean dissipation rate, $\nu \langle \ww ^2 \rangle$, and (c) the mean $6^{th}$ moment of the vorticity, $\langle \ww^{6} \rangle^{1/3}$, for runs 1-5, at fixed $\beta = 90^\circ$ and increasing values of $Re_\Gamma$.
The vertical dashed line in (b) and (c) corresponds
to $t = 23.2$, which is the last time shown in Fig.~\ref{fig:end_recon_b1}(c),
and is also very close to the peak dissipation rate. The dashed (respectively full) lines correspond to runs at low (respectively high) resolutions. 
}
\label{fig:Egy_Enst_b1}
\end{center}
\end{figure}
%%%%%%%%%%%%%%%%%%%%%%%%%%%%%%%%%%%%%%%%%%%%%%%%%%%%%%%%%%%%%%%

\paragraph*{Evolution of global quantities}
A quantitative measure of the production of small scales during the 
reconnection process is 
shown in Fig.~\ref{fig:Egy_Enst_b1} for the interaction of initially perpendicular vortex tubes at several Reynolds numbers. Over the entire period of the simulation, the mean kinetic energy rate, shown in Fig.~\ref{fig:Egy_Enst_b1}(a), decays by less than $20\%$,
despite the rapid increase in the mean dissipation rate, shown in 
Fig.~\ref{fig:Egy_Enst_b1}(b).
For each run, the initial mean dissipation rate decreases with a scaling of $\sim 1/Re_\Gamma$ as
the Reynolds number increases. It reaches a peak value at a time that is 
essentially independent of $Re_\Gamma$. We note that the peak dissipation time $t_{peak}$ will approximately coincide with the time the small-scale vortices are most energetic, so we can use this time to study the resulting small-scale structure.

Notably, the height of the
peak in the dissipation rate does not vary significantly as a function of the Reynolds number.
Further information on the generation of motion at
small scales is provided by higher moments of the vorticity distribution.
Specifically, Fig.~\ref{fig:Egy_Enst_b1}(c) shows the $6^{th}$ moment
of the vorticity, taken to power $1/3$. The $6^{th}$ moment is defined as:
\begin{equation}
\langle \ww^{2n} \rangle= \frac{1}{V} \int_V d^3 \xx ( \ww^2 )^n ~~\mbox{where} ~n=3
\end{equation} 
We chose to show a moment of finite order of the vorticity distribution, 
rather than the maximum of the vorticity, which corresponds to the limit
$n \rightarrow \infty$, as the latter is far more sensitive to finite
resolution effects. 

Since all runs share the same initial condition, the plots of $\langle \ww^6 \rangle^{1/3}$ all start at the same initial value at $0$. 
The peaks of the curves reach increasingly higher values with increasing $Re_\Gamma$; for the  $Re_\Gamma = 5000$ case the maximum is approximately $35$ times greater than the initial value. This trend reflects
the strong amplification of vorticity that occurs during reconnection as the cores contact and break down to fine scales.
Similar results are obtained with values of other moments of 
$\langle \ww^{2n} \rangle^{1/n}$, with $n = 2$ and $4$. As expected, the 
peak amplification for these moments grows with the order $n$.
Note that on all plots in Fig.~\ref{fig:Egy_Enst_b1}, the results of 
the runs at low resolution (with $N_l^3$ Fourier modes) are shown as
a dashed lines, whereas the runs at higher resolution (with $N_h^3$ modes,
$N_l$ and $N_h$ both given in Table~\ref{tab:1}) are represented by
solid lines. The deviations between the two resolutions are small in the peak
regions, even at the highest Reynolds number considered here. This
gives us confidence in our numerical results. 
However, at $Re_{\Gamma} = 5000$, the values of $\langle \ww^6 \rangle^{1/3}$ at different resolutions diverge
at later times (for $t \gtrsim 25$), which we interpret as a consequence
of the amplification of small differences in the numerical integration of
such a dynamical system with many degrees of freedom and underlying chaotic dynamics.

%%%%%%%%%%%%%%%%%%%%%%%%%%%%%%%%%%%%%%%%%%%%%%%%%%%%%%%%%%%%%%%
%% Figure 7 kinetic energy, dissipation, and 6th moment of vorticity for varying orientation angle
\begin{figure}[H]
\begin{center}
\subfigure[]{
\includegraphics[width=5.17cm]{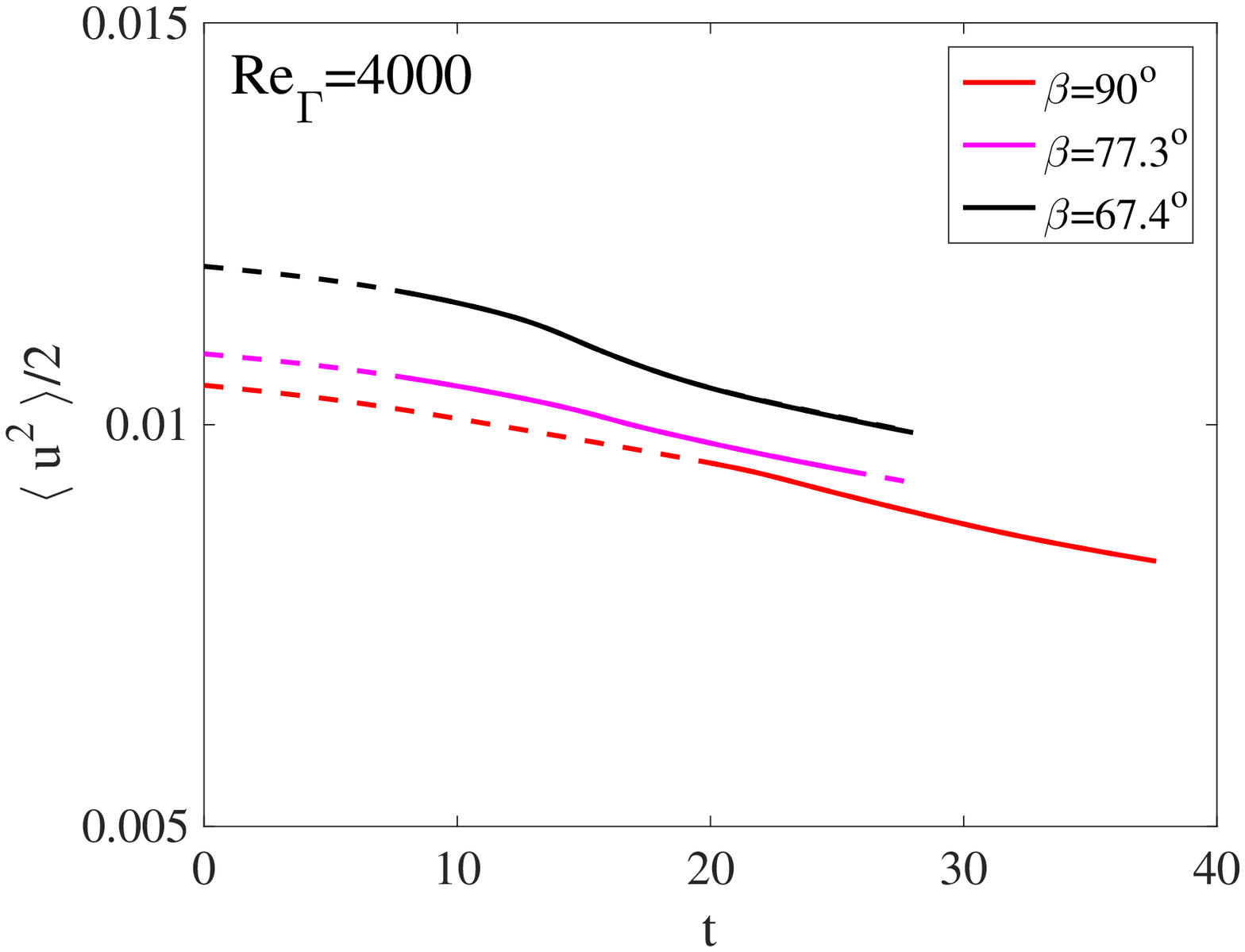}
}
\subfigure[]{
\includegraphics[width=5.17cm]{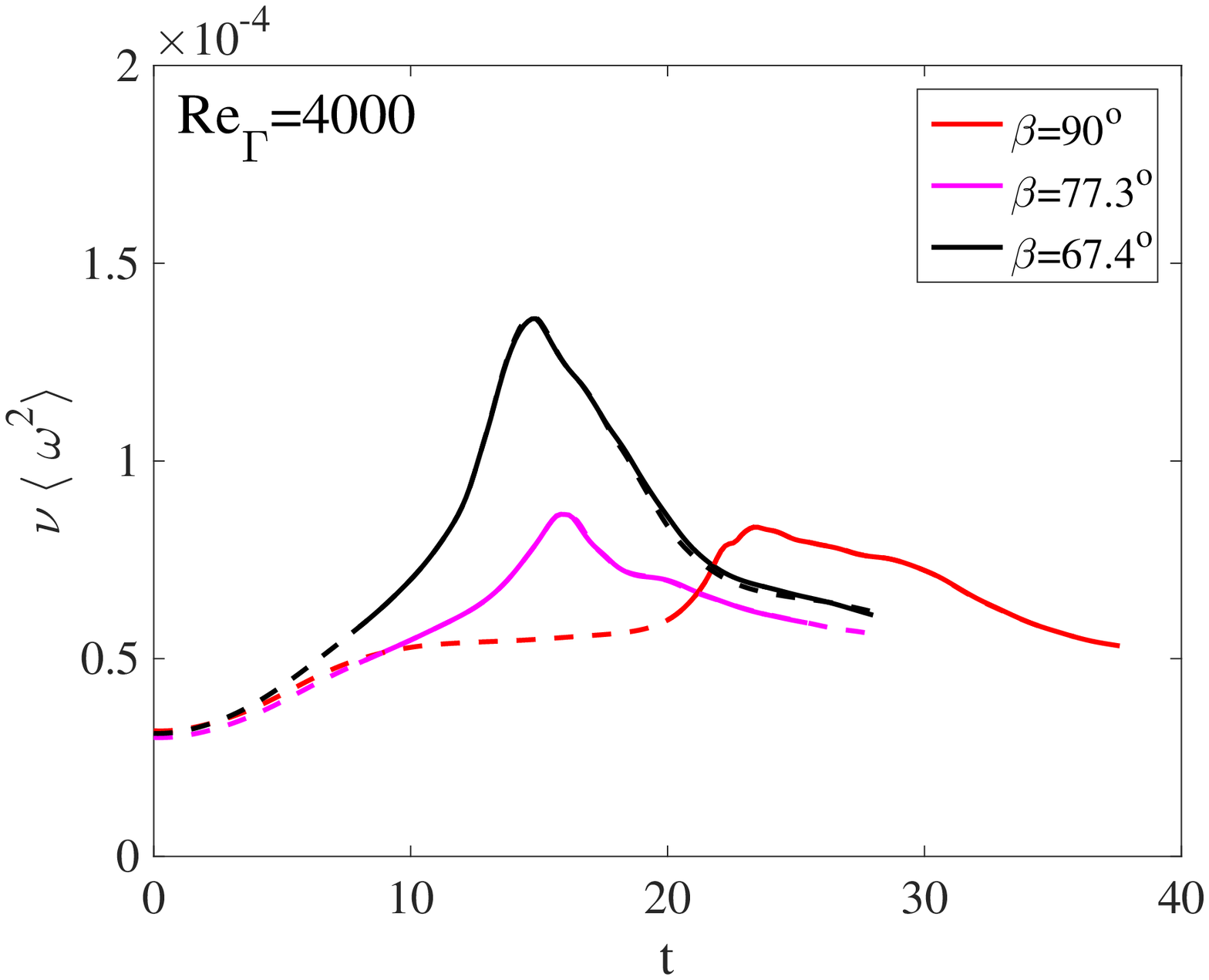}
}
\subfigure[]{
\includegraphics[width=5.17cm]{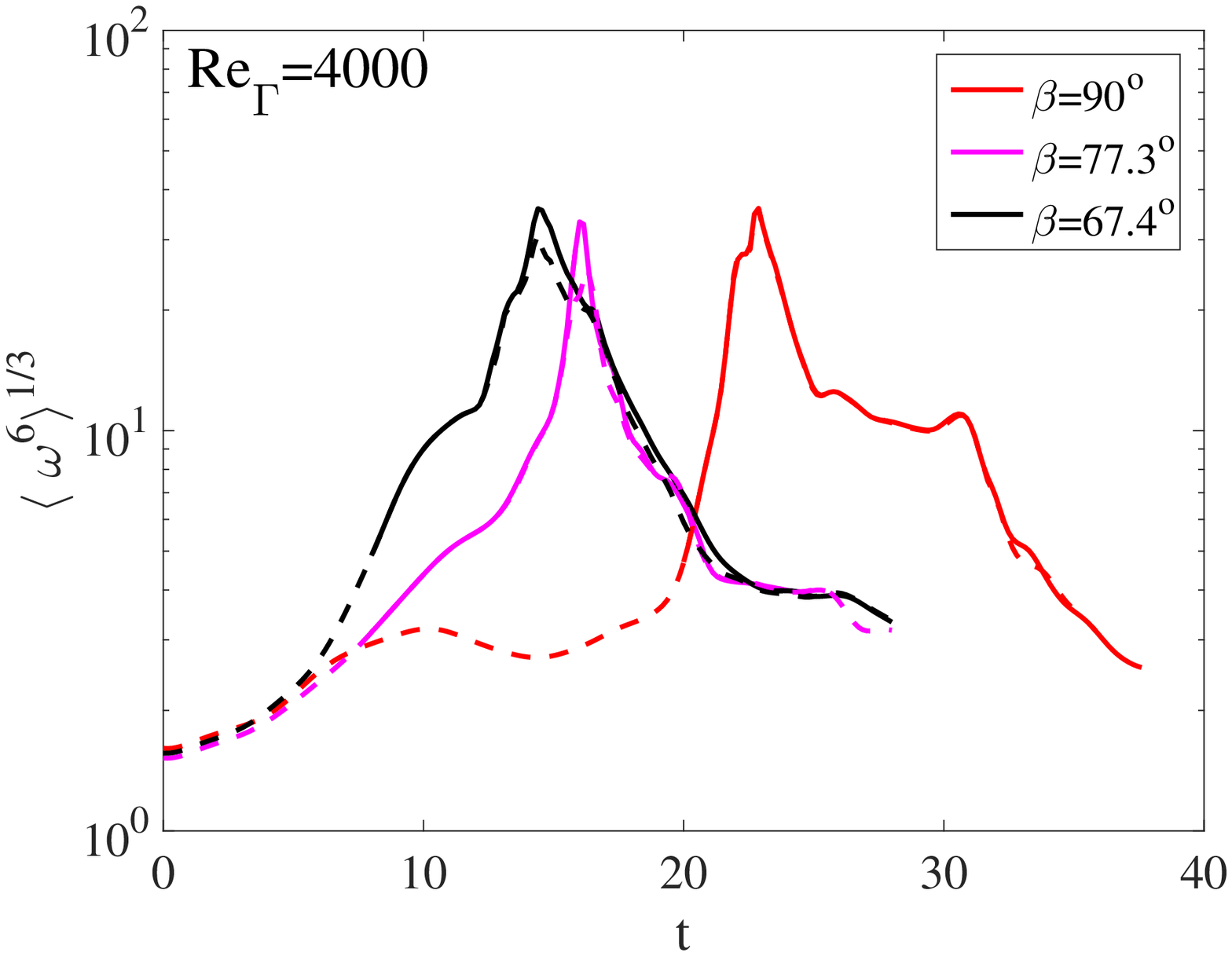}
}
\caption{
Global dynamics for vortex tubes at varying initial orientation angles.
The evolution of (a) the mean kinetic energy rate, $\langle \ve u ^2 \rangle /2$, (b) the mean dissipation rate, $\nu \langle \ww ^2 \rangle$, and (c) the mean $6^{th}$ moment of the vorticity, $\langle \ww^{6} \rangle^{1/3}$, for runs 3 ($\beta = 90^\circ$, $b=1$), 6 ($\beta \approx 77.3^\circ$, $b=5/4$), and 7 ($\beta \approx 67.4^\circ$, $b=3/2$), at fixed $Re_\Gamma = 4000$. The local pairing of the vortex tubes leads, in all of these cases, to the formation of vortex sheets.
}
\label{fig:egy_enst_b_lt_1.5}
\end{center}
\end{figure}
%%%%%%%%%%%%%%%%%%%%%%%%%%%%%%%%%%%%%%%%%%%%%%%%%%%%%%%%%%%%%%%

\paragraph*{Evolution at $\beta \gtrsim 67.4^\circ$ ($ b \le 3/2$)}
The evolution of the global quantities for runs 3, 6 and 7, all corresponding
to $Re_\Gamma = 4000$, and $ 67.4^\circ \le \beta \le 90^\circ$ ($1 \le b \le 3/2$) is shown in 
Fig.~\ref{fig:egy_enst_b_lt_1.5}. The initial value of the mean kinetic energy
slightly increases with $b$, as shown in Fig.~\ref{fig:egy_enst_b_lt_1.5}(a), and only decays by about 20\% throughout the whole run, as it was the case at $\beta = 90^\circ$ (compare with Fig.~\ref{fig:Egy_Enst_b1}). The main differences between the runs is indicated by the evolution of the mean dissipation rate and of $\langle \ww^6 \rangle^{1/3}$, as shown in Fig.~\ref{fig:egy_enst_b_lt_1.5}(b-c). Namely, as $\beta$ decreases, the time required for these plots to reach their respective maxima also decreases. Visualization studies, comparable to what has been done in the case
$\beta = 90^\circ$, show that the peaks in Fig.~\ref{fig:egy_enst_b_lt_1.5}(b-c)
correspond to the time at which the vortices reconnect and the
tubes change topology. Notably, the reconnection dynamics look comparable 
when $\beta = 90^\circ$ and $\beta \approx 77.3^\circ$ ($b = 5/4$), with the prominent formation of vortex sheets as
illustrated in Fig.~\ref{fig:inter_b1}.  
The major difference between these two cases comes from the time at which
the peaks develop, see 
Fig.~\ref{fig:egy_enst_b_lt_1.5}(b-c).
This difference can be qualitatively 
understood from the observation that when $\beta$ decreases, the vortex tubes
are initially closer to being antiparallel, and it therefore takes less 
time for them to locally align in an antiparallel manner and initiate the reconnection process. 
As explained earlier, this early phase can be captured with the Biot-Savart dynamics.

Contrary to the peaks of $\langle \ww^6 \rangle^{1/3}$
%for $\beta = 90^\circ$ and $\beta = 77.3^\circ$, see 
shown in Fig.~\ref{fig:egy_enst_b_lt_1.5}(c), which are approximately constant,
the peak energy dissipation rate is approximately twice as large for the configuration where $\beta \approx 67.4^\circ$ ($b = 3/2$) than for $\beta = 90^\circ$ ($b = 1$) or $\beta \approx 77.3^\circ$ ($b = 5/4$).
Visualization in the former case also reveals the transient presence of vortex
sheets when the tubes come together. The evolution, however, qualitatively
differs from what was shown in Fig.~\ref{fig:inter_b1}, as the sheets do
not stay close to one another. We view this regime as a transition towards
the dynamics occurring at smaller values of $\beta$, which will be discussed
in the following subsection.

\subsection{Evolution at $\beta \le 53.1^\circ$ ($b \ge 2$): short wavelength instability }
\label{subsec:ell_instab}

\begin{figure}[tb]
\begin{center}
\subfigure[]{
\includegraphics[width=5.17cm]{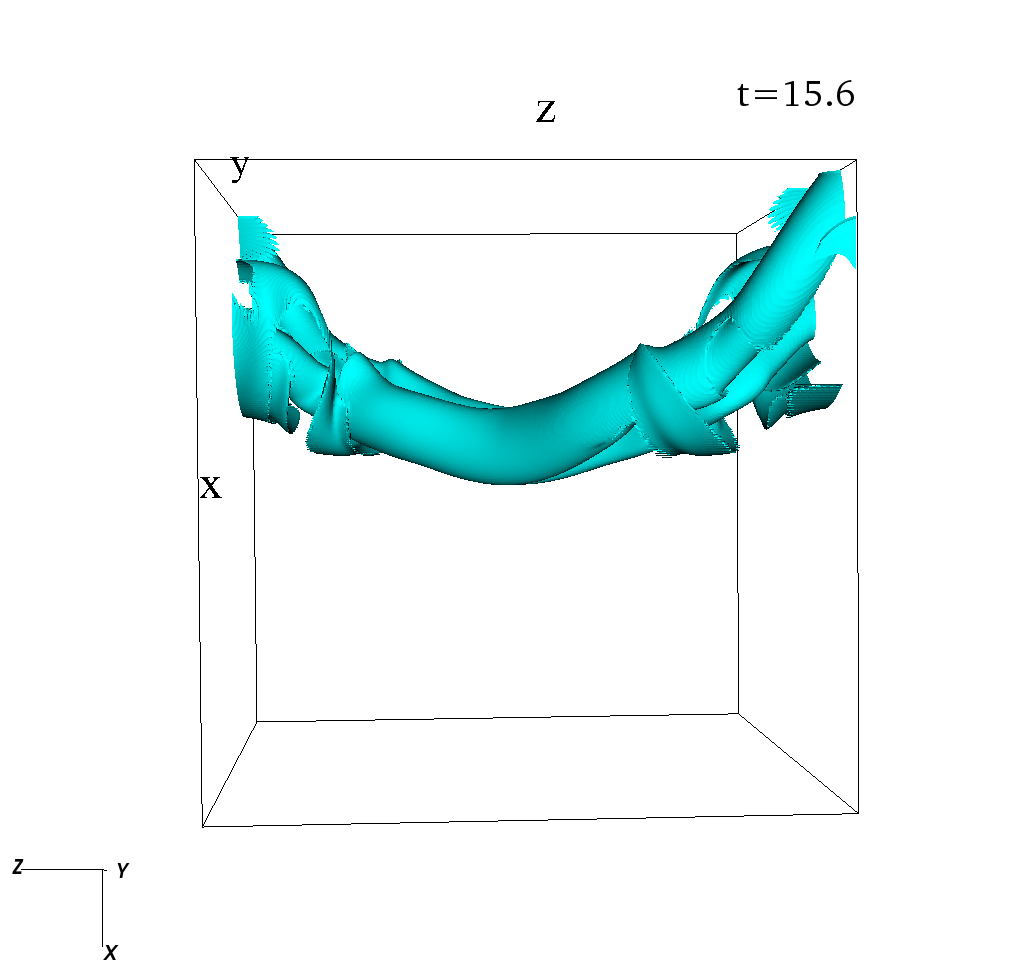}
}
\subfigure[]{
\includegraphics[width=5.17cm]{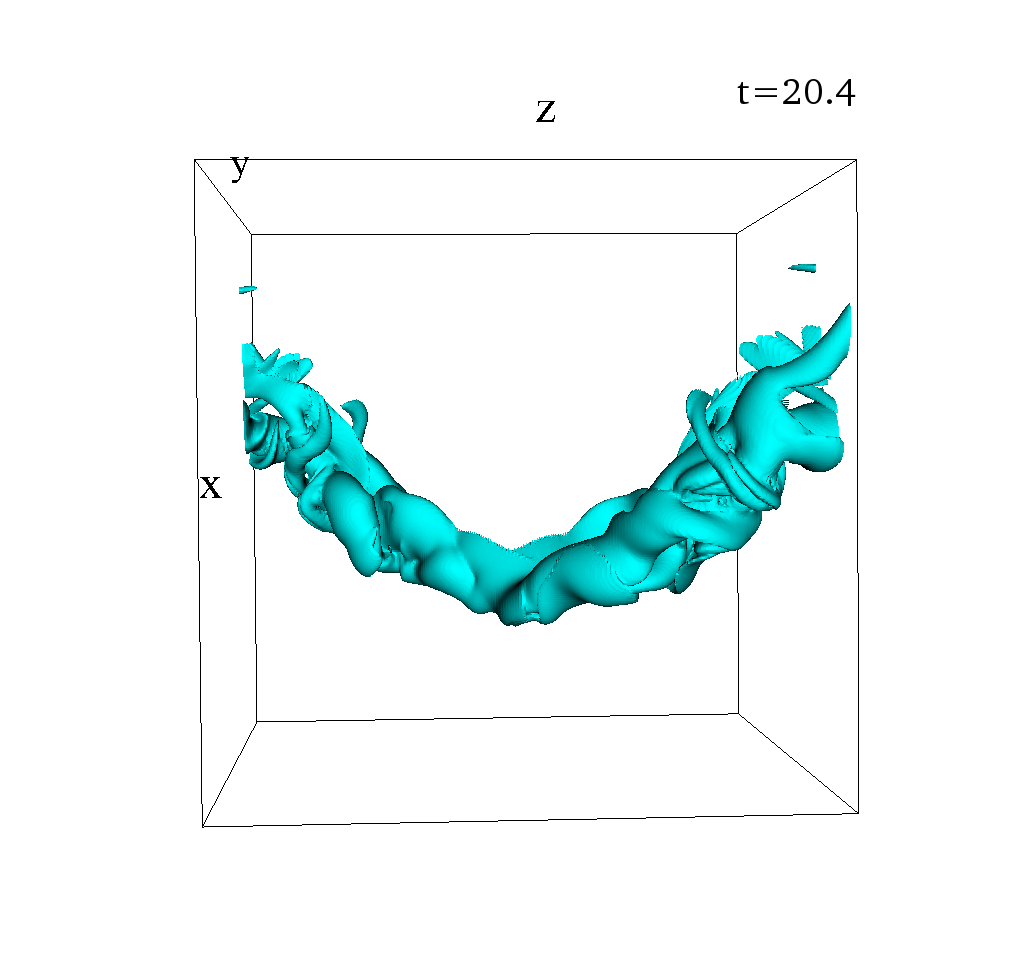}
}
\subfigure[]{
\includegraphics[width=5.17cm]{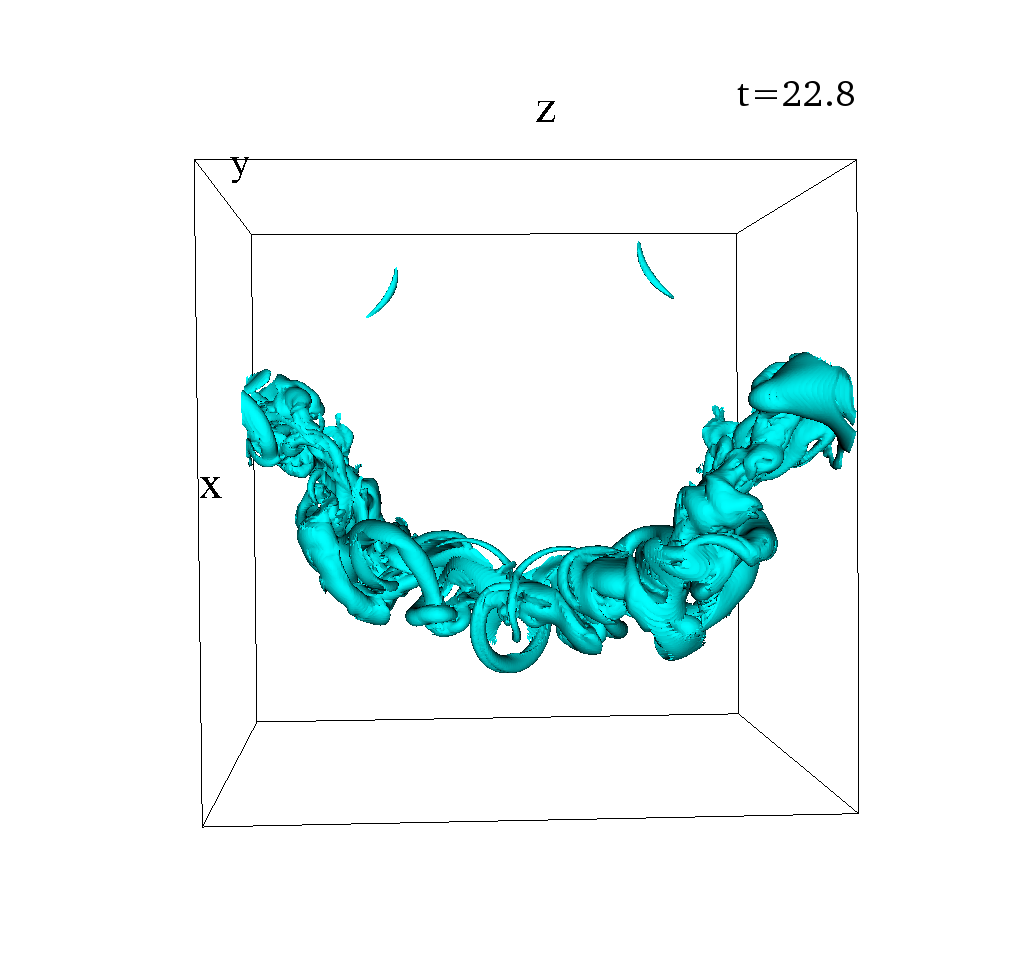}
}
\subfigure[]{
\includegraphics[width=5.17cm]{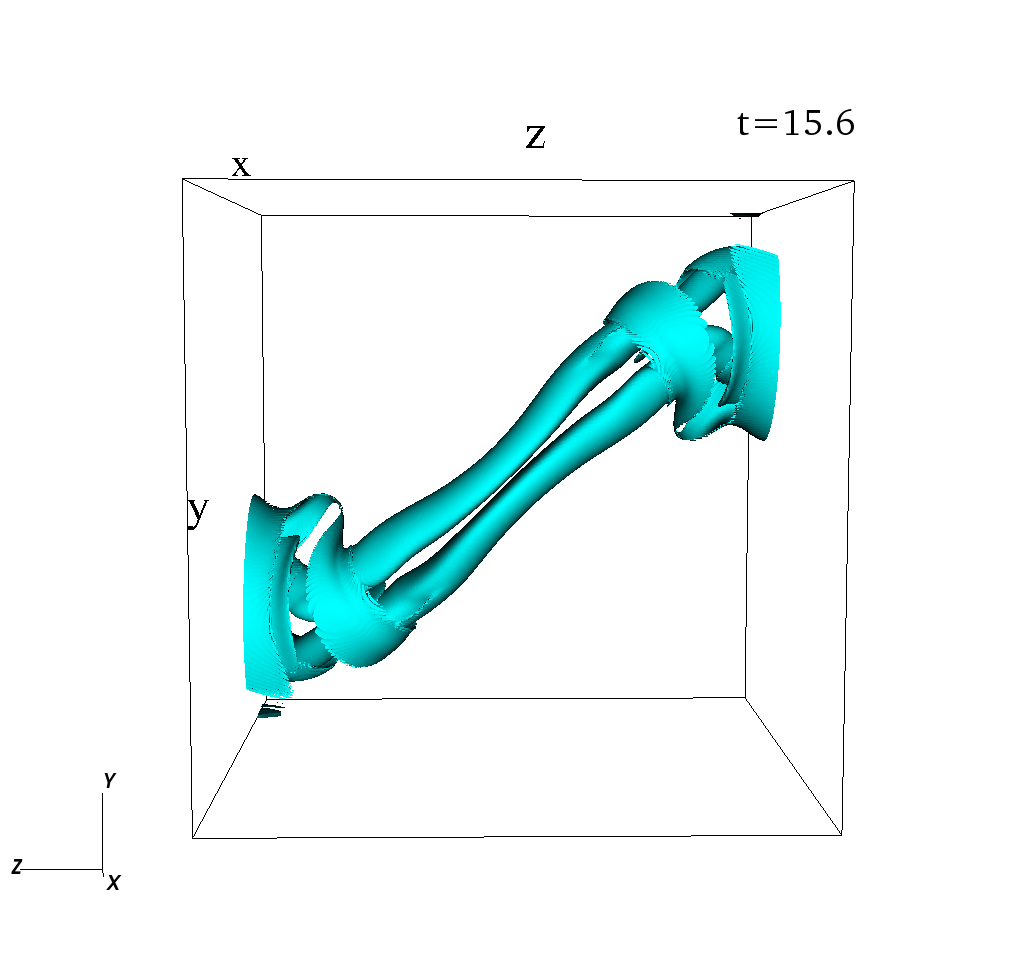}
}
\subfigure[]{
\includegraphics[width=5.17cm]{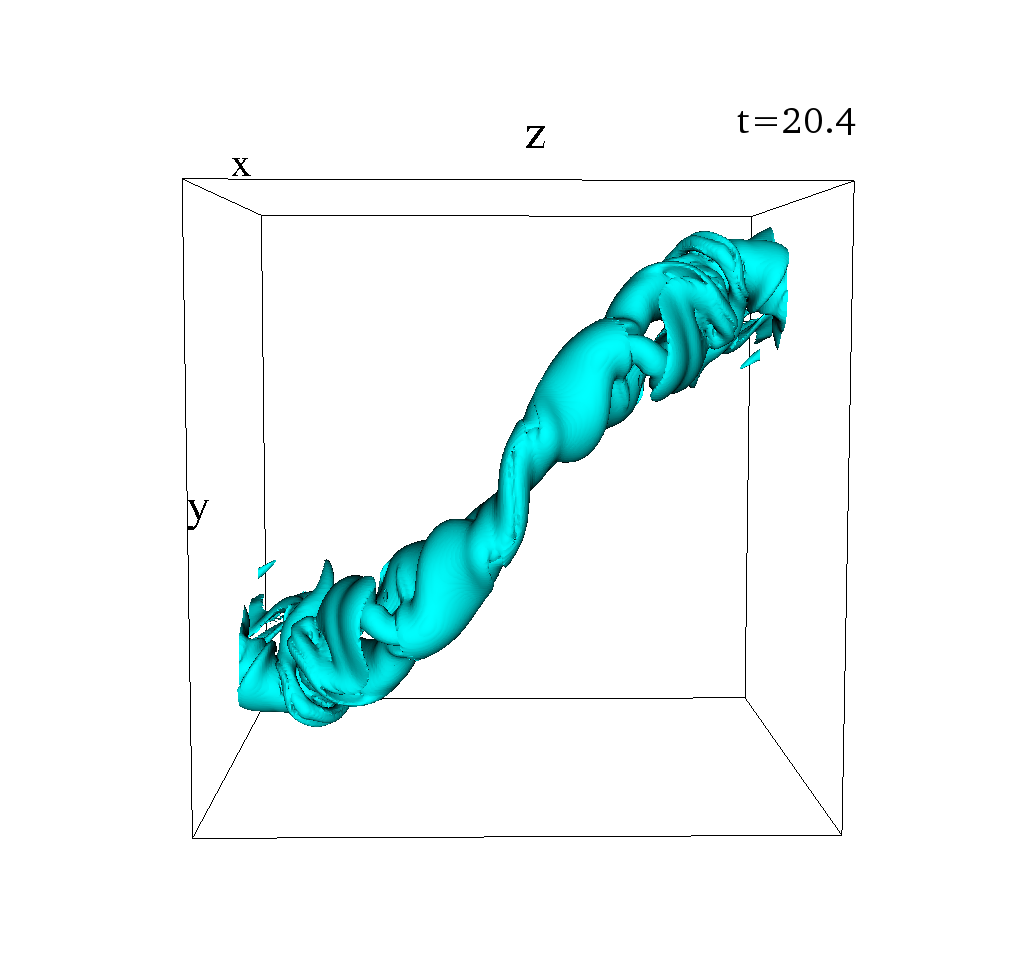}
}
\subfigure[]{
\includegraphics[width=5.17cm]{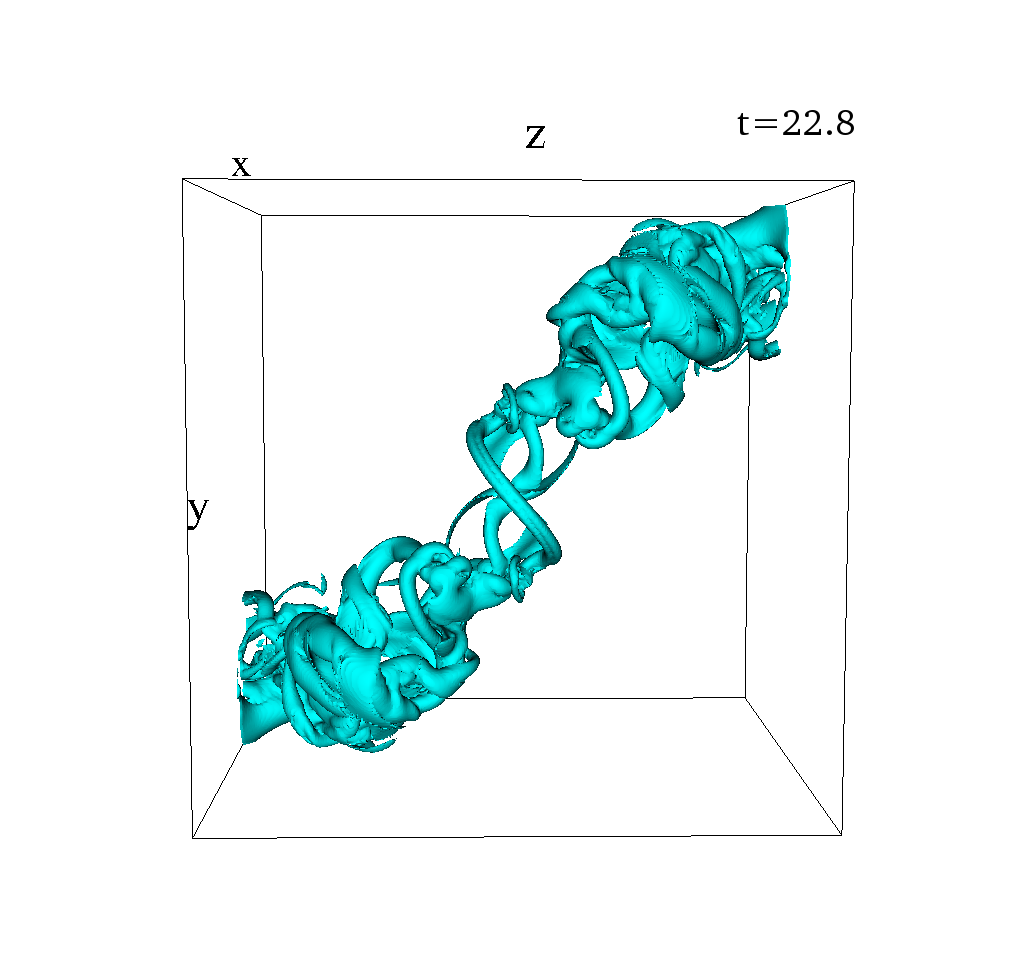}
}
\caption{Interaction and breakdown of nearly antiparallel vortex tubes.
The evolution of the vorticity magnitude isosurface for run 11 ($\beta \approx 28.1^\circ$, $b = 4$, $Re_{\Gamma} = 4000$). Only a cubic subdomain, $[ -2.6, 2.6]^3$, surrounding the regions where the vortices interact is shown. The isosurfaces of the vorticity field are shown at $t = 15.6$ (left), $t = 20.4$ (center) and $t = 22.8$ (right); the upper row shows the top view; and the bottom row shows the front view. The 
thresholds are $ \omega_{thr} = 2.5$ ($t = 15.6$), $3.2$ ($t = 20.4$)
and $3.6$ ($t = 22.8$). The latest time shown corresponds to the peak 
dissipation rate.
}
\label{fig:inter_b4}
\end{center}
\end{figure}

As the initial condition is varied to increase the alignment
of the tubes (i.e.~$\beta\to0$), the interactions close to the reconnection 
event become very different to what was shown for initially perpendicular tubes in Fig.~\ref{fig:inter_b1}. This is illustrated by Fig.~\ref{fig:inter_b4}, which
shows the development of the interaction between the vortex tubes
for run 11 with $\beta \approx 28.1^\circ$ ($b = 4$) and $Re_\Gamma = 4000$. Because the angle between the two tubes is much smaller than the initially perpendicular case ($b = 1$), the two tubes align, overlap, and interact over a significantly larger extent
which is much larger than the initial vortex core size. 
This is clearly visible in the left column of Fig.~\ref{fig:inter_b4}, as the extent of the vortices in 
the $z$-direction is much larger than in the $x$- and $y$- directions.
In fact, as the flow evolves from $t = 15.6$ (left column (a)) to $t = 20.4$
(central column), the vortices develop an instability over a wavelength
comparable to the size of the core. As the instability further evolves, 
many small-scale vortices develop (c.f.~the right panel of Fig.~\ref{fig:inter_b4}) through the mutual stretching and straining of perpendicular vortices which is mediated by the elliptical instability \cite{McKeown:2020}.

%%%%%%%%%%%%%%%%%%%%%%%%%%%%%%%%%%%%%%%%%%%%%%%%%%%%%%%%%%%%%%%
%% Figure 8 Visualization of vorticity evolution for shallow angled interaction (b=4) case
%%%%%%%%%%%%%%%%%%%%%%%%%%%%%%%%%%%%%%%%%%%%%%%%%%%%%%%%%%%%%%%

To characterize the development of the instability after $t \gtrsim 15.6$
and identify the means by which interacting tubes develop small-scale flow structures, clearly visible at $t = 20.4$ (in the middle column of
Fig.~\ref{fig:inter_b4}), we tracked the centerlines of the tubes. 
Recall that the tubes are initially located on the $(x,z)$-plane 
and separated by a distance $d$ in the $y$-direction, see Fig.~\ref{fig:scheme}.
For small enough values of $\beta$, as the flow evolves, the tubes move primarily
in the $y$- and $x$-directions as the vortex axis is almost parallel to 
the $z$ direction. At each value of $z$ along the axis of the tubes, 
we separate the $y$-domain into two subdomains, $D_\pm$, corresponding to the the two tubes, as clearly visible from the front view at $T = 6.13$ in Fig.~\ref{fig:inter_b4}. 
In practice, this is done by computing the integral
of $\ww^4$ over $x$: $\zeta(y,z) = \int dx' \ww^4(x',y,z)$ and by identifying, at each position $z$,
the value of $y$ that separates the upper and lower part of the tube.
We then determined the $x$-location of the 
centroids at each value of $z$ by computing the moments 
$ x_\pm = \bigl( \int_{D_\pm} d x dy \, \ww^4 x \bigr)/\bigl(\int_{D_\pm} dx dy \, \ww^4 \bigr) $, with a similar definition for $y_\pm$. We note that this way of defining the location of the centerlines fails as the two tubes begin to interpenetrate, as shown at $t = 20.4$ in Fig.~\ref{fig:inter_b4}.

\begin{figure}[tb]
\begin{center}
\subfigure[]{
\includegraphics[width=5.9cm]{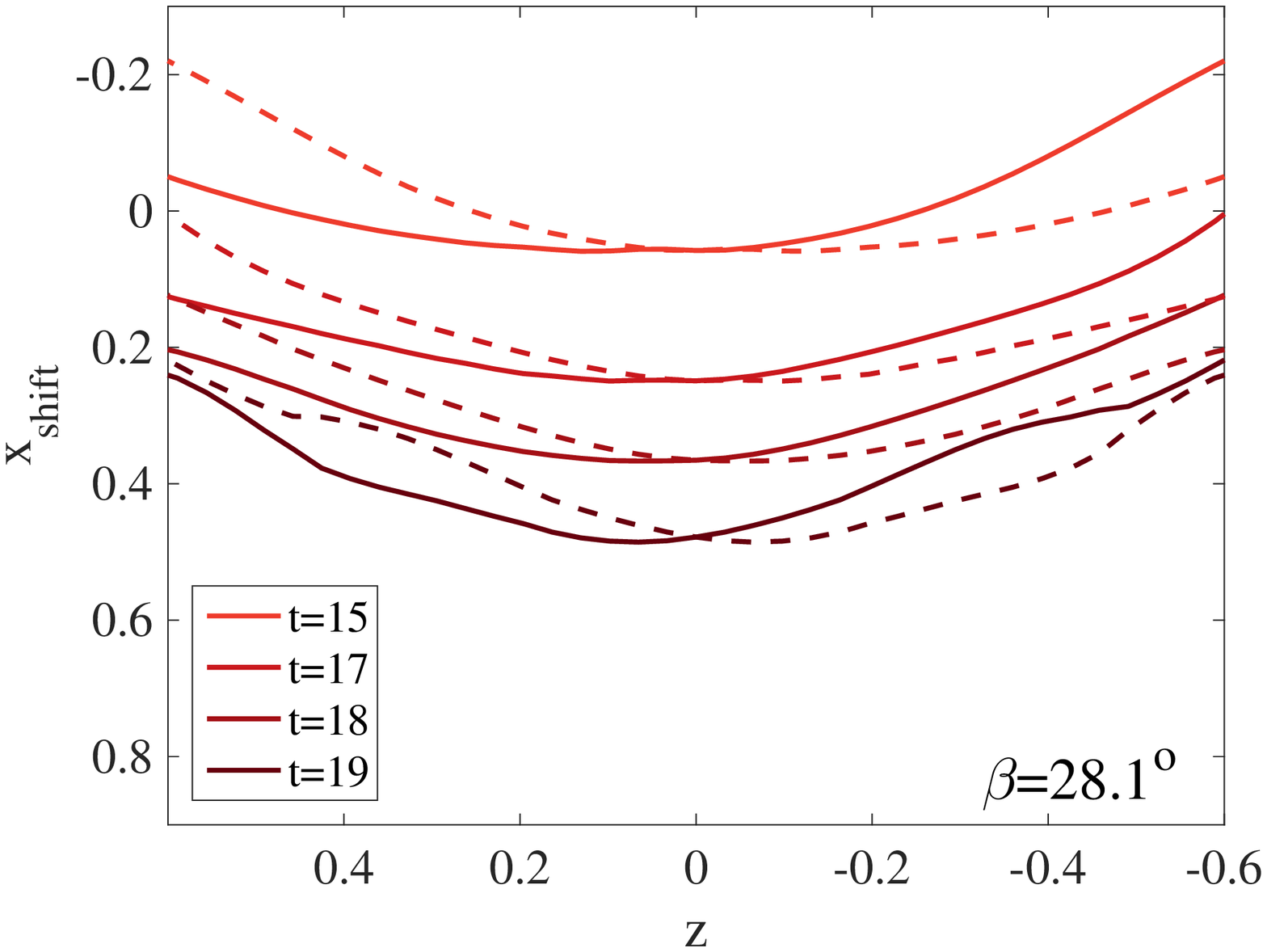}
}
\subfigure[]{
\includegraphics[width=5.9cm]{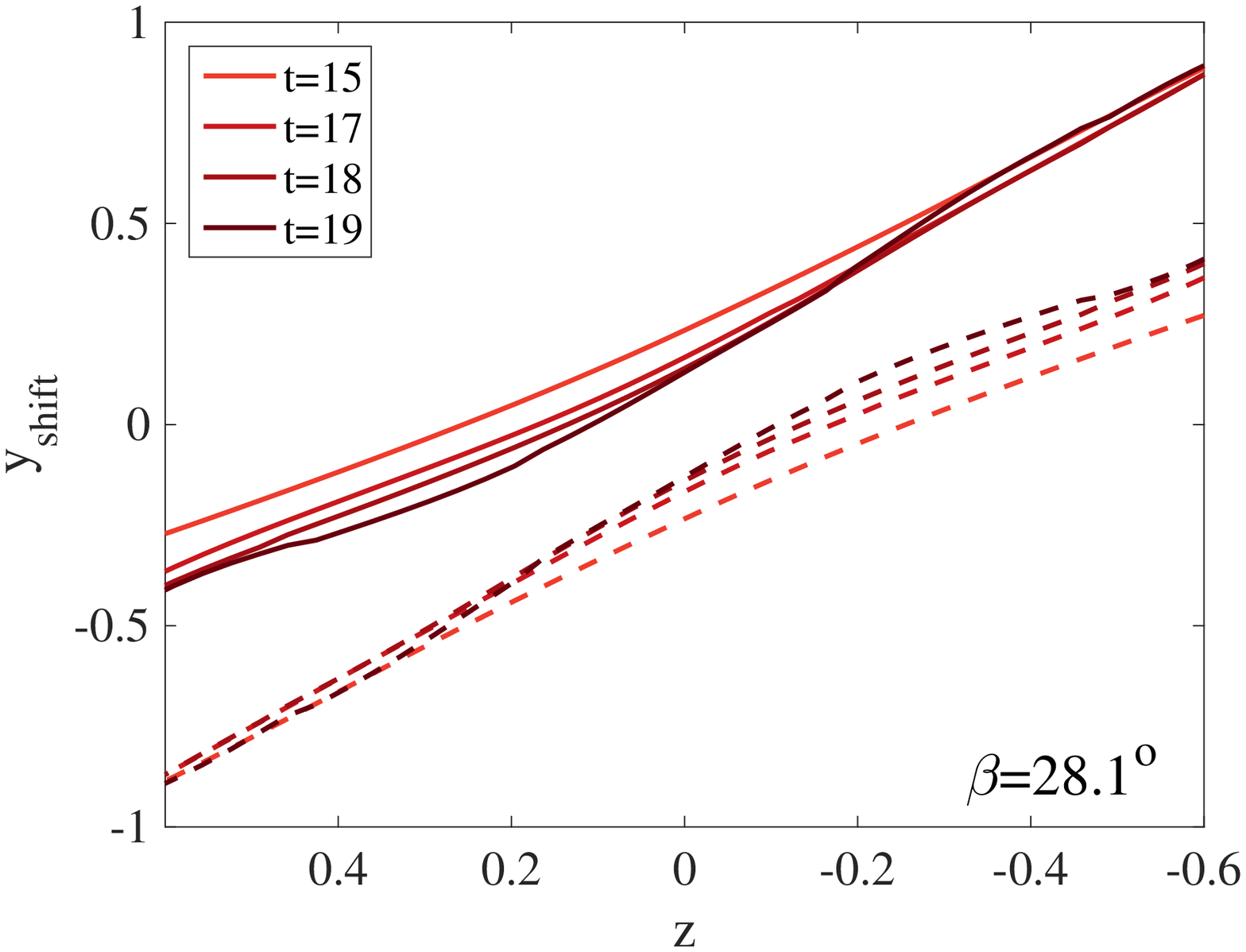}
}
\caption{Vortex centerline trajectories for nearly antiparallel vortex tubes for run 11 ($\beta \approx 28.1^\circ$, $b=4$, $Re_{\Gamma} = 4000$). (a) Top view and (b) Side view. The solid lines correspond to the upper vortex and the dashed line correspond to the lower vortex.
The centerlines were extracted by computing the moments of $\ww^4 x$, $y$ and $z$.
in a portion of the domain, including the central region where the vortices
interact.}
% ($- 2.6 \le x , y \le 2.6$, $ -4.25 \le z \le 4.25$). }
\label{fig:centroid_b4}
\end{center}
\end{figure}

Fig.~\ref{fig:centroid_b4} shows the top (a) and front views (b) of the vortex tube centerlines; the full lines indicate the upper vortex, and the dashed 
lines indicate the lower vortex.
The evolution of the centerline locations suggests the development of an 
instability with a wavelength comparable to the core radii . 
In terms of distinguishing between which vortex instability drives the growth of this perturbation--either
the Crow instability~\cite{Crow:1970} or the elliptic instability~\cite{Leweke:1998}-- we are faced with the difficulty that the symmetry of the modes is not as clear as in the case of two initially parallel vortex tubes ($\beta = 0$), as found in \cite{McKeown:2020}. 
This can be easily understood, given the
relatively small size of the cores in interaction and the constraints
on either side of the region of interaction. Nonetheless, the small
distance between the cores, and the latest stage of the development shown
in Fig.~\ref{fig:centroid_b4}, where perpendicular filaments are formed, suggests the prevalence as time progresses
of a symmetry that corresponds more to the elliptic instability, than 
to the Crow instability, reminiscent of what was observed 
in~\cite{McKeown:2018,McKeown:2020}. 

The results shown in this subsection contrast sharply with those
shown in Section~\ref{subsec:sheets} for $\beta \gtrsim 67.4^\circ$ ($b < 3/2 $). In fact, we can 
distinguish between two very different qualitative behaviors. For $\beta \gtrsim 67.4^\circ$,  ($ b < 3/2$), 
the interaction of the vortex tubes leads to the localized formation of thin vortex sheets that are limited to the interaction zone. 
In contrast, for $\beta \lesssim 53.1^\circ$, ($b \ge 2$), the interaction of the two vortex tubes leads to an interaction similar to that shown in Fig.~\ref{fig:inter_b4}, where perpendicular, fine-scale filaments arise throughout large areas of the domain. We stress the qualitative resemblance of the production of small-scales in this figure with the results shown in~\cite{McKeown:2018,McKeown:2020}.  
We recall that the flow corresponding the run 7 ($\beta \approx 67.4^\circ$, $b = 3/2$) does lead to the development of sheets,
but the interaction mechanism ultimately differs from those shown in 
Fig.~\ref{fig:inter_b1} and in Fig.~\ref{fig:inter_b4} because even if small-scale perpendicular filaments arise they come in small numbers and do not interact with each other significantly.

%%%%%%%%%%%%%%%%%%%%%%%%%%%%%%%%%%%%%%%%%%%%%%%%%%%%%%%%%%%%%%%
%% Figure 9 Vortex Centerline trajectories for b=4 case
%%%%%%%%%%%%%%%%%%%%%%%%%%%%%%%%%%%%%%%%%%%%%%%%%%%%%%%%%%%%%%%

\begin{figure}[tb]
\begin{center}
\subfigure[]{
\includegraphics[width=5.17cm]{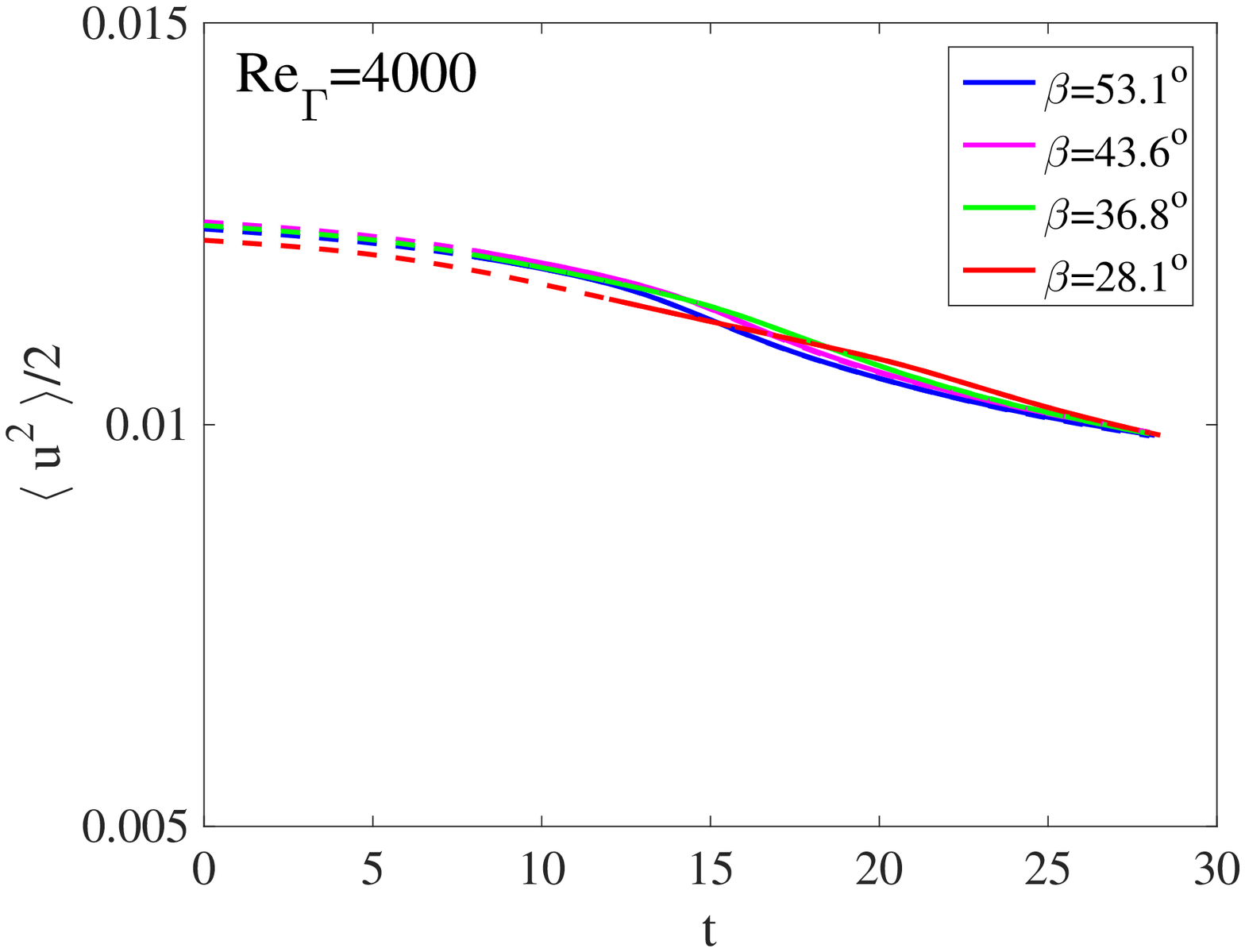}
}
\subfigure[]{
\includegraphics[width=5.17cm]{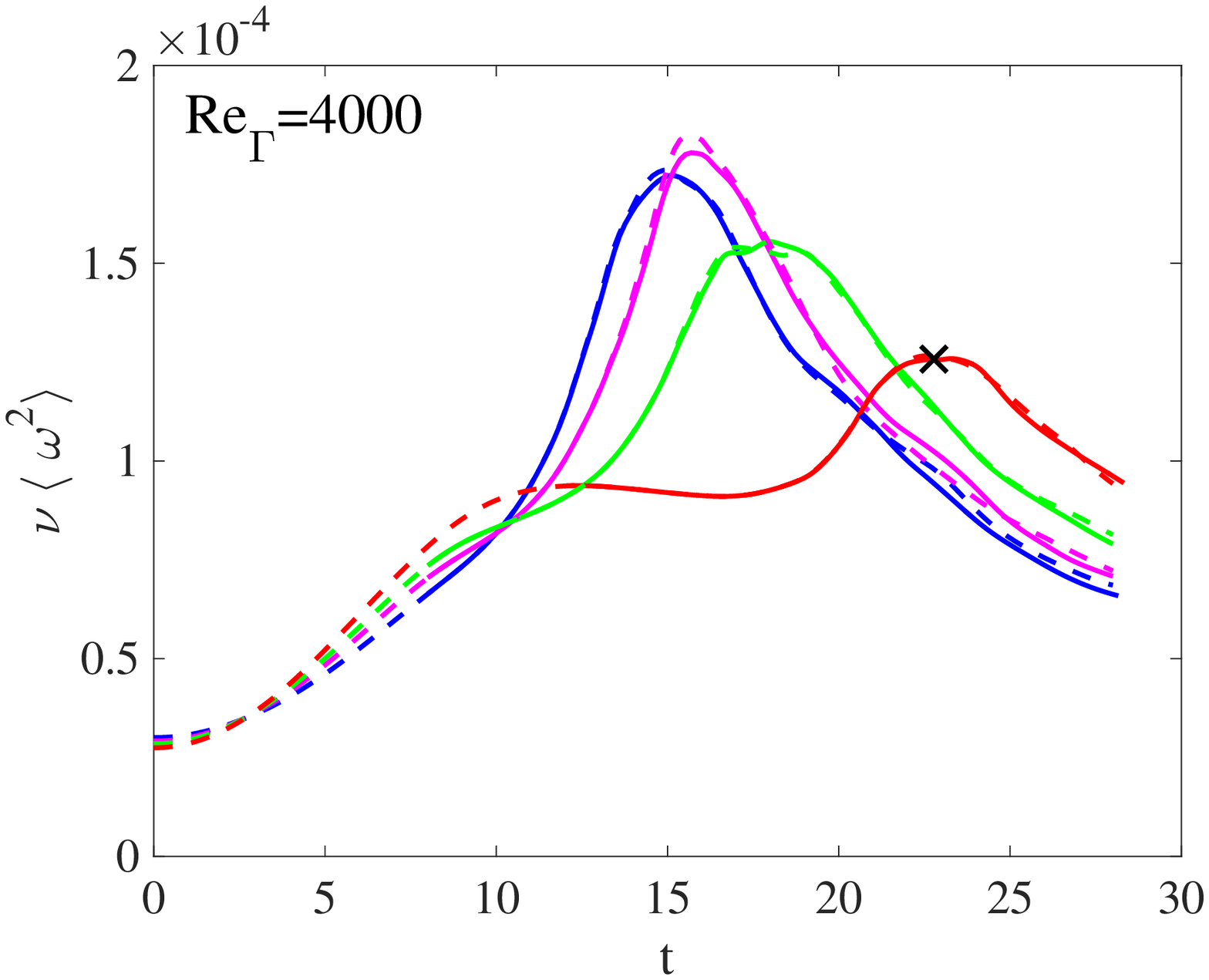}
}
\subfigure[]{
\includegraphics[width=5.17cm]{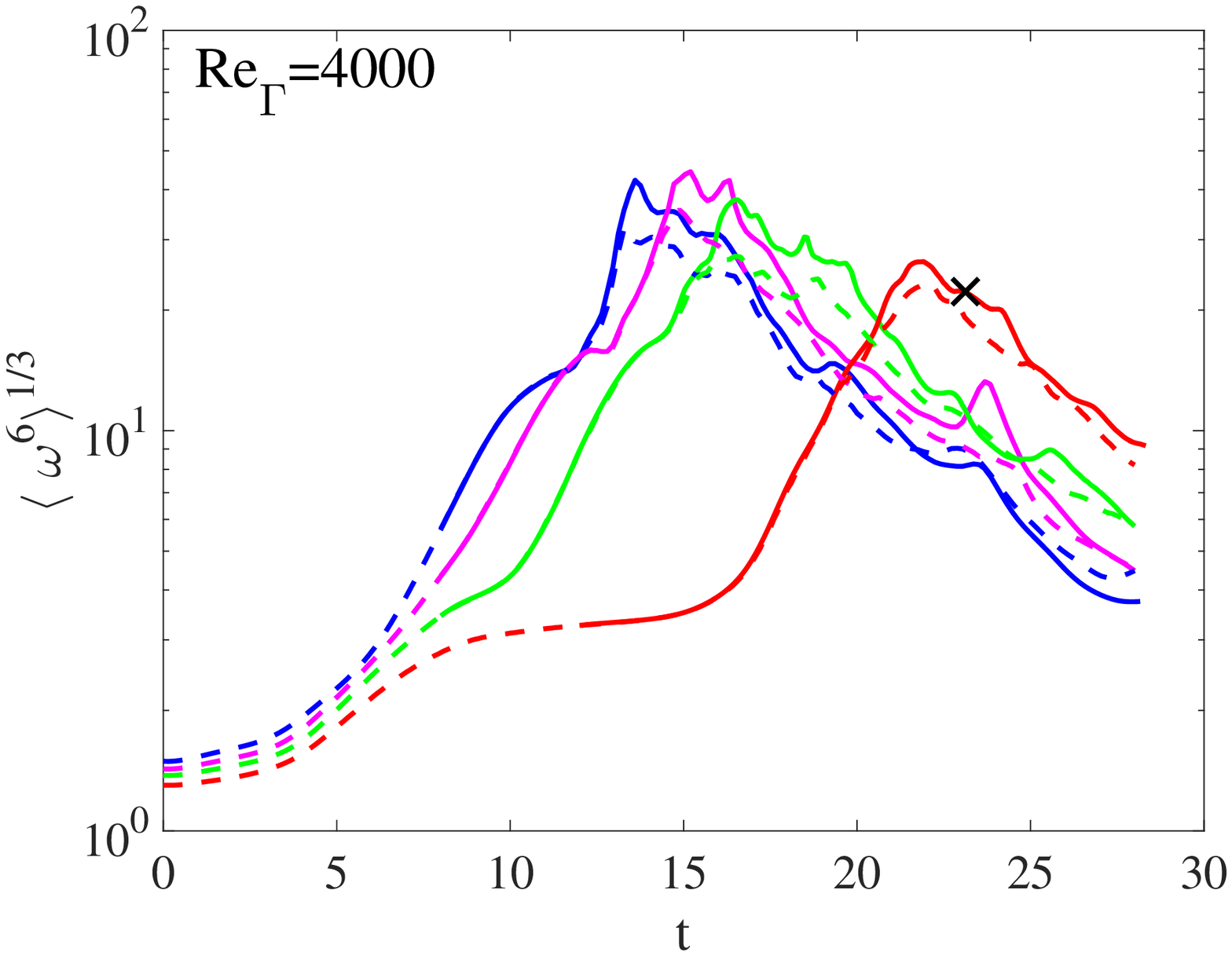}
}
\caption{
Global dynamics for vortex tubes initially oriented at shallow angles.
The evolution of (a) the mean kinetic energy rate, $\langle \ve u ^2 \rangle /2$, (b) the mean dissipation rate, $\nu \langle \ww ^2 \rangle$, and (c) the mean $6^{th}$ moment of the vorticity, $\langle \ww^{6} \rangle^{1/3}$, for runs 8 ($\beta \approx 53.1^\circ$, $b=2$), 9 ($\beta \approx 43.6^\circ$, $b=5/2$), 10 ($\beta \approx 36.8^\circ$, $b=3$), and 11 ($\beta \approx 28.1^\circ$,  $b=4$), at fixed $Re_\Gamma = 4000$.
}
\label{fig:egy_enst_b_ge_2}
\end{center}
\end{figure}

Fig.\ref{fig:egy_enst_b_ge_2} shows the time-dependence of the
kinetic energy of the runs (panel a), the dissipation rate (b), and the $6^{th}$
moment of vorticity, $\langle \ww^6 \rangle^{1/3}$ for runs 8-11, where 
$\beta \le 53.1^\circ$ ($b \ge 2$) and $Re_{\Gamma} = 4000$.
We have indicated by a cross in Fig.~\ref{fig:egy_enst_b_ge_2}(b-c)
the latest time corresponding to the visualization in Fig.~\ref{fig:inter_b4}(c), which approximately coincides with the peak dissipation rate. As was the case for the runs at $\beta \gtrsim 67.4^\circ$ ($b \le 3/2$), see Fig.~\ref{fig:egy_enst_b_lt_1.5}),
the time of the peak dissipation, $t_{peak}$, varies with $\beta$. 
Fig.~\ref{fig:egy_enst_b_ge_2}b shows that $t_{peak}$ increases when $\beta$ decreses. 
It should be kept in mind that the time at which the interaction occurs
is a consequence of the pairing process, which depends on the precise geometry
of the problem, and more specifically, on the angle $\beta$ between the vortex tubes. 
As the tubes become more parallel,
it takes a longer time for the interaction between the tubes to initiate.
In the limit of perfectly antiparallel filaments, this time becomes  
the time necessary for instabilities to grow, as observed 
in~\cite{McKeown:2020}. It depends on the amount of
noise initially, and it can be  much longer than the time $t \approx 25$
for run 11 ($\beta \approx 28.1^\circ$, $b = 4$).
In fact, we checked that the triggering
of the elliptic instability, leading to the strong interaction between
two antiparallel vortex tubes, is delayed when decreasing the amplitude of the
noise added to the solution. This is consistent with the intuitive notion 
that the interaction leading to turbulence starts with an exponential growth 
of a small perturbation of the two initially antiparallel vortex tubes.
 
Interestingly, we also notice that the value of the peak dissipation rate tends
to decrease when $\beta$ decreases, for $\beta \lesssim 43.6^\circ$ ($b \gtrsim 5/2$). A similar trend is also observed in the 
$6^{th}$ moment, see Fig.~\ref{fig:egy_enst_b_ge_2}(c) as well as for the fourth and $8^{th}$
moments (not shown). 
%This suggests that the generation of small-scales is most intense at $\beta \approx 45^\circ$. 
Contrary to runs 
at higher values of $\beta$, we observe a stronger difference between 
the runs at low resolution (with $N_l$ Fourier modes, shown as dashed lines), 
and the runs at a higher resolution (with $N_h$ Fourier modes, shown as
full lines). This indicates stronger resolution requirements for these runs.

%%%%%%%%%%%%%%%%%%%%%%%%%%%%%%%%%%%%%%%%%%%%%%%%%%%%%%%%%%%%%%%
%% Figure 10 Global dynamics for nearly antiparallel vortex tubes

\subsection{Discussion}
\label{subsec:discus}

Whereas the interaction between vortex tubes always leads to reconnection i.e. to 
a change of topology of the vortex lines, the mechanisms involved  
when the initial conditions are close to anti-parallel ($\beta \lesssim 33.7^\circ$ or $b > 3/2$, see subsection~\ref{subsec:ell_instab}), 
appear to qualitatively differ from what is observed when the vortices are closer to being perpendicular ($\beta \gtrsim 33.7^\circ$ or $b < 3/2$) 
as discussed in subsection~\ref{subsec:sheets}. 
The dynamics observed in the former case 
are very reminiscent of what was observed during the interaction of 
two initially antiparallel vortex tubes~\cite{McKeown:2018,McKeown:2020}. 
The qualitative similarity between reconnection when $b = 4$, occurring through the annihilation 
of a large fraction of the two locally antiparallel tubes, clearly shown in Fig.~\ref{fig:recon_b4} and
\ref{fig:inter_b4} and the dynamics resulting from the collision
between two vortex rings~\cite{McKeown:2020} is an important aspect of our work. 

\begin{figure}[tb]
\begin{center}
\subfigure[]{
\includegraphics[width=5.17cm]{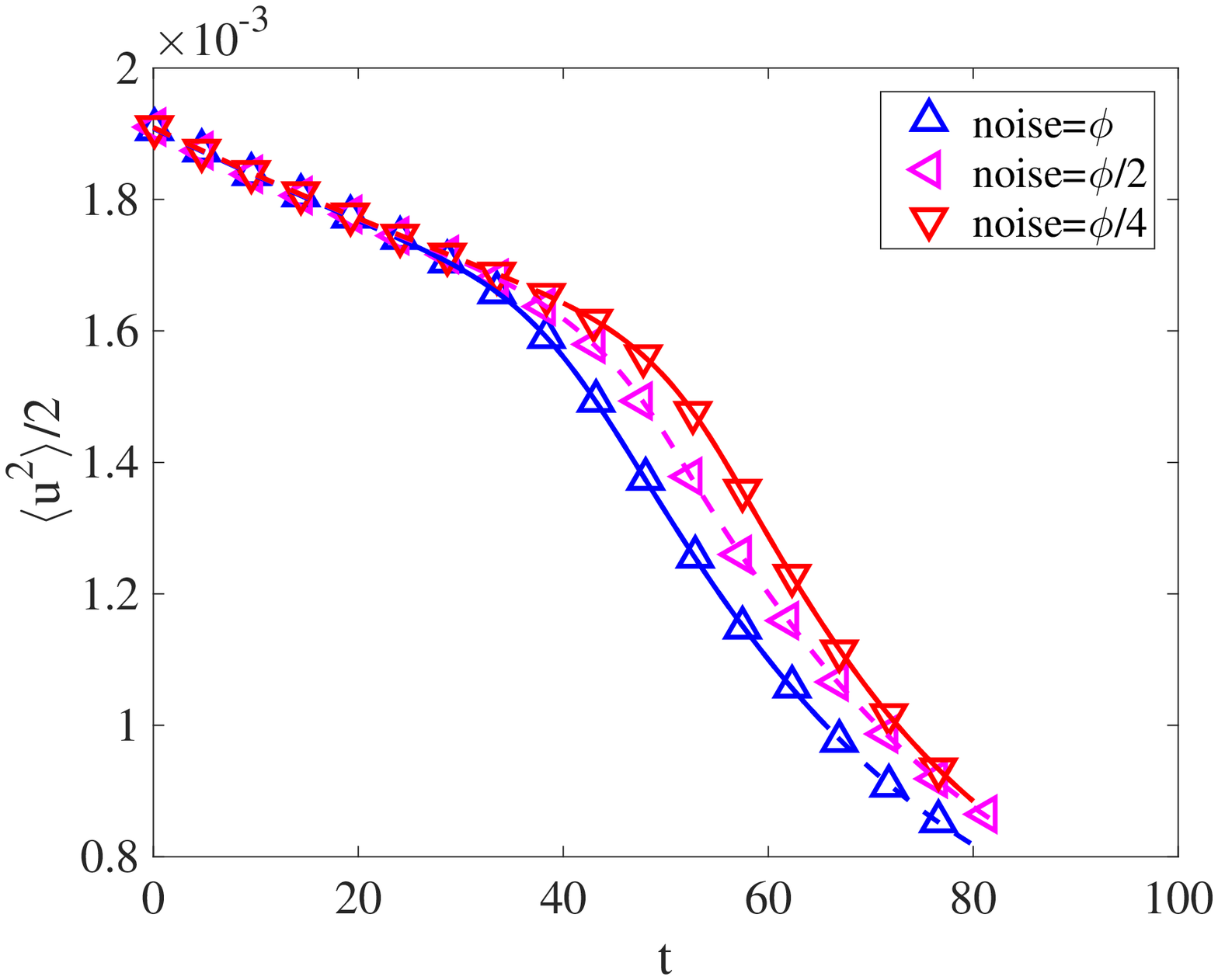}
}
\subfigure[]{
\includegraphics[width=5.17cm]{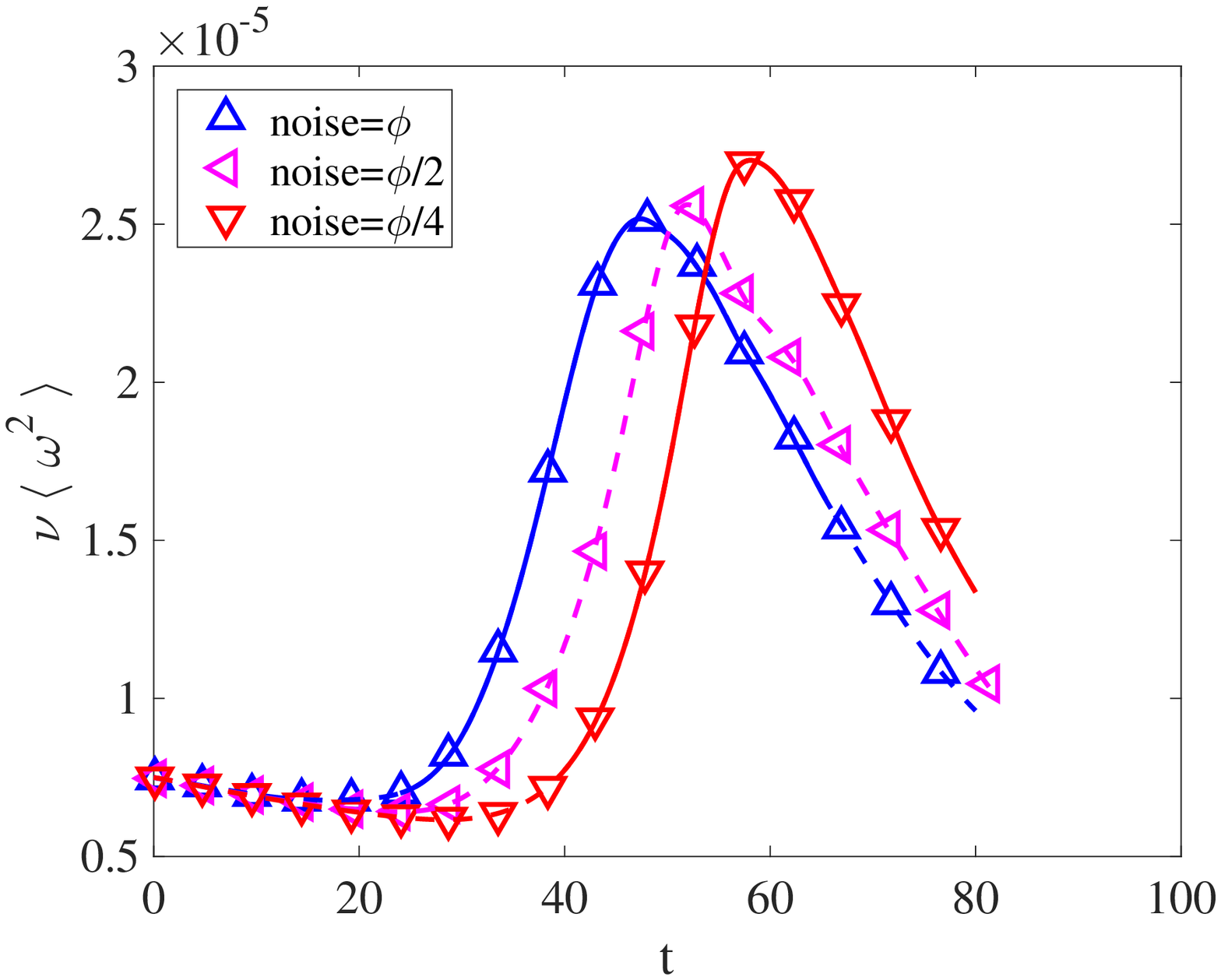}
}
\subfigure[]{
\includegraphics[width=5.17cm]{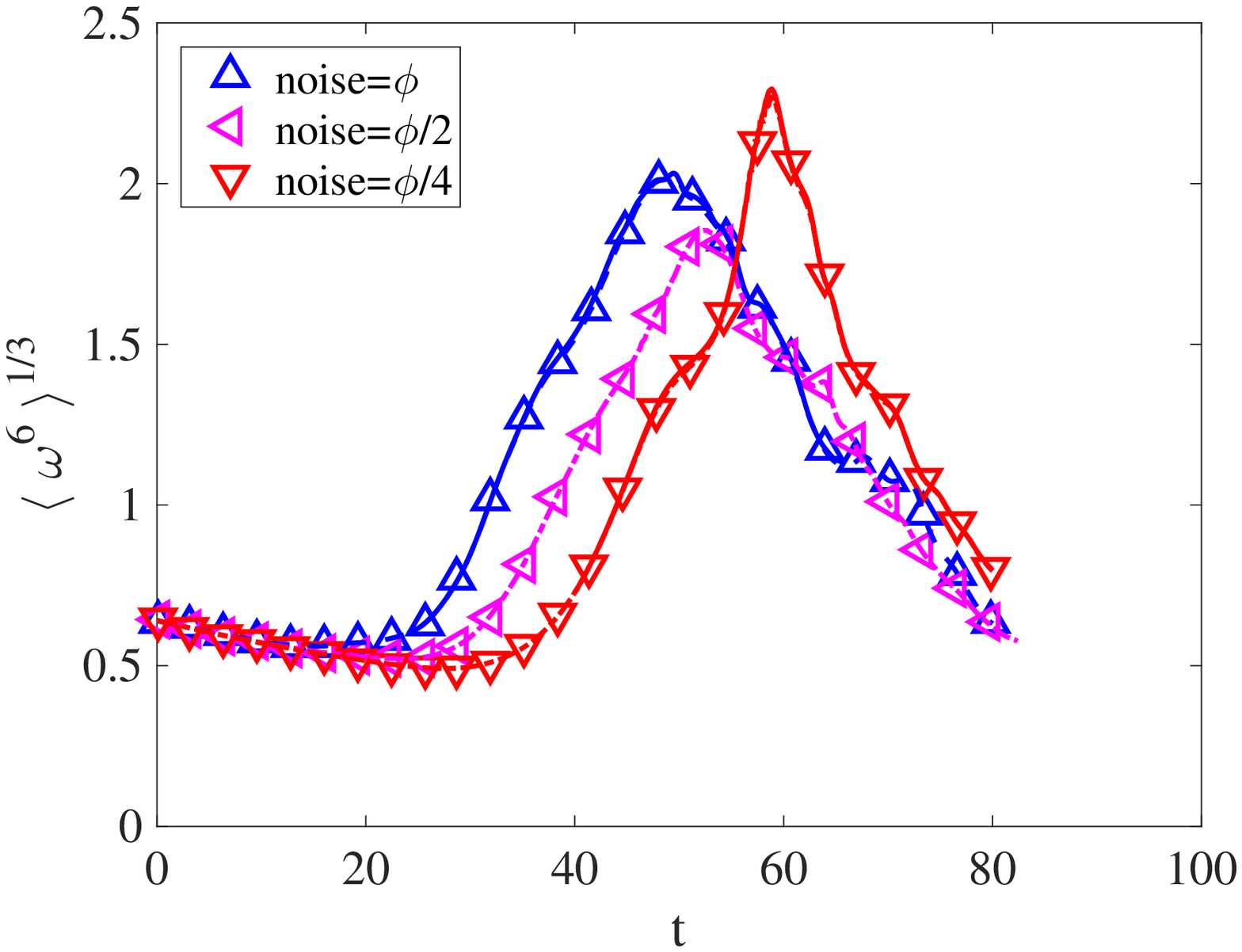}
}
\caption{Global dynamics for initially antiparallel vortex tubes with various noise levels.
The evolution of (a) the mean kinetic energy rate, $\langle \ve u ^2 \rangle /2$, (b) the mean dissipation rate, $\nu \langle \ww ^2 \rangle$, and (c) the mean $6^{th}$ moment of the vorticity, $\langle \ww^{6} \rangle^{1/3}$, for run 12 with varying noise amplitudes where $\beta = 0^\circ$ 
and $Re_\Gamma = 4000$.
}
\label{fig:egy_enstr_par}
\end{center}
\end{figure}

This configuration of two antiparallel tubes corresponds formally to
$\beta\to 0^\circ$ ($b \to \infty$). In fact, the behaviors of the mean kinetic energy, dissipation rate, and $6^{th}$ moment of vorticity in runs with initially weakly perturbed antiparallel vortex tubes, see Fig.~\ref{fig:egy_enstr_par}, are very comparable to that shown for $\beta=28.1^\circ$. The main difference is that the time at which the violent interaction leads to the breakdown of the vortex tubes and generation of fine-scale flow structures begins at later times, compared to what is seen in Fig.~\ref{fig:egy_enst_b_ge_2}(b-c). 
Furthermore, the time at which the interaction occurs depends on the level of noise. This can be clearly seen in Fig.~\ref{fig:egy_enstr_par}(b-c), which compares 3 simulations with three different noise levels, obtained from a weak level (upward triangle symbols), half its value (left pointing triangles) and a quarter of its
value (downward triangles). The time where the dissipation rate peaks clearly depends on the destabilization of the initial noise amplitude: we observe a logarithmic dependence of this time, consistent with the intuitive notion that
the first stage of the interaction comes from the exponential growth of an unstable perturbation, through the elliptic instability.

%%%%%%%%%%%%%%%%%%%%%%%%%%%%%%%%%%%%%%%%%%%%%%%%%%%%%%%%%%%%%%%
%% Figure 11 Global dynamics for antiparallel vortex tubes with different noise levels
%%%%%%%%%%%%%%%%%%%%%%%%%%%%%%%%%%%%%%%%%%%%%%%%%%%%%%%%%%%%%%%

This signals a clear qualitative difference between this configuration, where the interaction between the tubes leads to the annihilation of increasingly larger parts of overlapping tubes, and that obtained for $\beta > 67.4^\circ$ ($b < 3/2$), where intense, extremely thin vortex sheets form before a reconnection event~\cite{Pumir:1987,Kerr:1993,Yao:2020a}. 
%This demonstrates that the process leading to the energy cascade is not universal but rather strongly depends on geometry. 
Both types of event lead to the formation of small scale structures, albeit through different dynamics.
We postulate that for sufficiently large Reynolds numbers and small values of $\beta$, reconnection is overtaken by the mutual annihilation of the two tubes through 
the elliptical instability. This is because the elliptical instability requires the strain to be  aligned along the vortex core to begin to act (hence it barely acts for $\beta = 90^\circ$), but its growth rate is much larger than that of the Crow instability that leads to reconnection \cite{McKeown:2020}. 
Even if the overall dynamics preceding reconnection appear to depend on the initial condition, the late stage of the interaction leads at high enough Reynolds numbers in all cases studied, to an intense generation of small-scales, plausibly through a cascade as demonstrated in the case of parallel tubes in~\cite{McKeown:2020}. It is tempting to postulate that this cascade through the generation of perpendicular, small-scale vortices, may in fact be universal~\cite{Brenner:2016,Goto:2017}, independently of the initial conditions.

\section{Summary and conclusion}

In this work, we have investigated the interaction between two
initially straight, counterrotating vortex tubes oriented at an angle $\beta$. We systematically varied $\beta$, hence the geometry of the initial flow configuration, and let the flow evolve. In all cases, we observe a change in topology of the vortex lines and the production of perpendicular small-scale vortices. 
The main result of our
study is that the dynamics which result in this outcome depend strongly on
the initial orientation of the interacting tubes. 

When the
tubes are initially almost perpendicular to each other ($\beta \approx \pi/2$), the
interacting vortex tubes locally contact where they overlap, flattening into a pair of intense, slender vortex sheets.  This resembles the classically studied reconnection,  
found in the case of a pair of straight vortex tubes with
a perturbation symmetric with respect to the plane separating
the two vortices~\cite{Pumir:1987,Kerr:1989,Pumir:1990,Kerr:1993,Shelley:1993,Hou:2006,Kerr:2013,Brenner:2016,Yao:2020a}, 
and which involves the flattening of the vortex tubes into intense, slender vortex sheets on either side of the symmetry plane.
The underlying symmetries in the flow appear to be consistent with the eigenmodes 
corresponding to the long-wavelength (Crow) instability \cite{Crow:1970},
and can be adequately captured in a flow with imposed symmetries. The breakdown of the tubes ultimately leads to abundant formation of small scales, see also~\cite{Yao:2020a}.

However, when the angle between the two filaments is initially 
acute, we observe a breakdown mechanism that is comparable to that of Refs.~\cite{McKeown:2018,McKeown:2020}, with the formation of abundant small-scale vortices perpendicular to the original tubes, which effectively transfers more energy to the small-scales. The anti-symmetric elliptical instability drives the dynamics in this case, resulting in a breaking of the underlying symmetries discussed above. Hence, this evolution cannot be captured by flows with an imposed symmetry.

It is interesting to contrast this study with the non-universal aspects of vortex reconnection
in superfluids, documented in~\cite{Villois:2017}, which rest on the precise geometry of the filament pair when the vortices reconnect. The nonuniversality 
documented in the present work rests on the dynamics of the vortex cores, which are responsible for the onset and dynamics of the elliptic instability. The Reynolds number also appears to be an important parameter in the reconnection of vortex tubes in 
classical fluids, a flow parameter without an analog in superfluids.

While the early stages of vortex interactions do not appear to lead
to a universal reconnection scenario, we observe that
the interaction between two vortex tubes leads in seemingly 
different ways, to a proliferation of small scale flow structures. In the problem
considered here, when $\beta \approx \pi/2$,
the small scales form after the sheets have annihilated, as observed 
in~\cite{Yao:2020a}. When the tubes are better aligned in an antiparallel manner, the formation of small-scale vortex structures via a cascade occurs as soon as the
vortices come together~\cite{McKeown:2020}. 
The idea that iterative mechanisms may lead to formation of a cascade
has been suggested for theoretical reasons~\cite{Tao:2016,Brenner:2016}.
Strong evidence for a cascade driven by a hydrodynamic instability, namely
the elliptic instability, has been presented in the interaction of two
vortex tubes. 
An interesting question for future work will be to understand whether
the mechanism leading to the proliferation of small-scale vortices is universal,
based on the physical mechanisms discussed in~\cite{McKeown:2020}.
The possibility that such a cascade scenario may lead to a singularity, as
already postulated~\cite{Tao:2016,Brenner:2016,Yao:2020b} also deserves
further attention.

\noindent {\it Acknowledgments:} R.O.M. thanks the Research Computing Data Core (RCDC) at the University of Houston for providing computing resources. This research was funded by the National Science Foundation through the Harvard Materials Research Science and Engineering Center DMR-2011754, and through the Division of Mathematical Sciences DMS-1715477. M.P.B. is an investigator of the Simons Foundation. A.P. acknowledges financial support from the IDEXLYON project (Contract ANR-16-IDEX-0005) under University of Lyon auspices, as well as from the project TILT from ANR. 

\bibliography{biblio}

\end{document}